\documentclass[a4paper,12pt]{article}
\usepackage[fencedCode,inlineFootnotes,citations,definitionLists,hashEnumerators,smartEllipses,pipeTables,tableCaptions,hybrid]{markdown}
\usepackage{verbatim}
\usepackage{listings}
\usepackage{upquote}
\usepackage{graphicx}
\usepackage[normalem]{ulem}

\usepackage[labelfont=bf]{caption}
\usepackage{float}
\usepackage{slashed}
\usepackage{stmaryrd}
\usepackage{subcaption}
\usepackage{caption}
\usepackage[utf8]{inputenc}     
\usepackage[T1]{fontenc} 
\usepackage{xparse}
\usepackage{multirow}
\usepackage[english]{babel}     
\usepackage{xcolor} 
\definecolor{linkcolor}{rgb}{0,0,0.6} 
\usepackage{hyperref}
\usepackage{amssymb}
\usepackage{soul}
\usepackage{geometry}
\usepackage{latexsym}
\usepackage{cite}
\usepackage{latexsym}
\usepackage{framed}
\usepackage{cite}
\usepackage{soul}
\usepackage{graphicx}
\usepackage{booktabs}
\usepackage{amssymb,amsmath,amsthm}
\usepackage{cleveref}

\usepackage{physics}
\usepackage{mathtools}
\usepackage{amsthm}
\usepackage{bbold}
\usepackage{multirow}
\usepackage{pdfpages}
\usepackage{chngcntr}

\usepackage{fancyvrb}

\newcommand{\marty}{\Verb+MARTY+}
\definecolor{pmssm_color}{HTML}{004334}
\definecolor{nmfv_color}{HTML}{803900}

\newcommand{\pmssm}[1]{\ensuremath{\color{pmssm_color}#1}}
\newcommand{\nmfv}[1]{\ensuremath{\color{nmfv_color}#1^*}}
\newcommand{\GeV}{\ensuremath{\ \mathrm{GeV}}}
\newcommand{\TeV}{\ensuremath{\ \mathrm{TeV}}}

\newcommand{\Gc}{\gamma_{ 5}}

\newcommand{\ds}{\displaystyle}

\renewcommand{\a}{\alpha}

\renewcommand{\d}{\delta}

\newcommand{\g}{\gamma}

\newcommand{\sw}{\sin^2{\theta_W}}
\newcommand{\as}{\alpha_{s}}
\newcommand{\bea}{\begin{align}}
\newcommand{\eea}{\end{align}}
\newcommand{\beq}{\begin{equation}}
\newcommand{\eeq}{\end{equation}}

\newcommand{\fr}{\frac}
\newcommand{\be}{\begin{equation}}
\newcommand{\ee}{\end{equation}}
\def\bsp#1\esp{\begin{split}#1\end{split}}
\def\bpm{\begin{pmatrix}}
\def\epm{\end{pmatrix}}
\newcommand {\superiso} {{\tt SuperIso}}

\newcommand {\softsusy} {{\tt SOFTSUSY}}

\numberwithin{equation}{section}

\begin{document}
\hfill {\tt  CERN-TH-2022-001} 

\def\thefootnote{\fnsymbol{footnote}}

\begin{center}
{\Large\bf{Flavour anomalies in supersymmetric scenarios\\[0.2cm]
with non-minimal flavour violation}}

\setlength{\textwidth}{11cm}
                    
\vspace{2.cm}
{\large\bf  
M.A. Boussejra$^{a,}$\footnote{email: boussejra@ipnl.in2p3.fr}, F.~Mahmoudi$^{a,b,}$\footnote{email: nazila@cern.ch},
G.~Uhlrich$^{a,c,}$\footnote{email: gregoire.uhlrich@unige.ch}
}
 
\vspace{1.cm}
{\em $^a$Universit\'e de Lyon, Universit\'e Claude Bernard Lyon 1, CNRS/IN2P3, \\
Institut de Physique des 2 Infinis de Lyon, UMR 5822, F-69622, Villeurbanne, France}\\[0.2cm]
{\em $^b$Theoretical Physics Department, CERN, CH-1211 Geneva 23, Switzerland}\\[0.2cm] 
{\em $^c$D\'epartement de physique nucl\'eaire et corpusculaire, Universit\'e de Gen\`eve,\\
CH-1211 Geneva, Switzerland} 

\end{center}

\renewcommand{\thefootnote}{\arabic{footnote}}
\setcounter{footnote}{0}

\vspace{1.5cm}
\thispagestyle{empty}
\centerline{\bf ABSTRACT}
\vspace{1.cm}

Motivated by tensions between experimental measurements and SM predictions in $b\to s \ell^+\ell^-$ transitions, we present the first study of non-minimal flavour-violating Minimal Supersymmetric Standard Model (MSSM) scenarios contributing to the relevant Wilson coefficients to address the observed anomalies using \superiso\ and \marty.
We calculate the full one-loop analytical contributions of the general MSSM to Wilson coefficients relevant for flavour anomalies, together with the anomalous muon magnetic dipole moment $(g-2)_\mu$.
We show that, after imposing theoretical constraints on the flavour-violating parameters we can find scenarios in agreement with the experimental measurements that can address at the same time the tensions in flavour observables and in $(g-2)_\mu$.

\clearpage

\section{Introduction}

In recent years, impressive progress has been achieved in studying and measuring semileptonic $B$ decays. In particular, neutral currents with $b \to s$ transitions offer a plethora of clean observables that have been under scrutiny as they present tensions with the Standard Model (SM) predictions.
The first tension, at the level of 3$\sigma$, was reported in 2013 in the measurement of angular observables related to $B \to K^* \mu^+\mu^-$ decay~\cite{LHCb:2013ghj}. Since then, similar tensions have been observed in several decays, such as $B\to K \mu^+\mu^-, B_s\to \phi \mu^+\mu^-$ and $\Lambda_b \to \Lambda \mu^+\mu^-$~\cite{LHCb:2014cxe,LHCb:2015tgy,LHCb:2020gog,LHCb:2021zwz}. 
In addition, LHCb measured lepton-flavour universality violating (LFUV) ratios $R_{K^{(*)}}={\rm BR}(B\to K^{(*)}\mu\mu)/{\rm BR}(B\to K^{(*)}ee)$, that are predicted very precisely in the SM, and confirmed the tension with the SM with about 3$\sigma$ significance for low dilepton mass squared ($q^2$)~\cite{LHCb:2017avl,LHCb:2021trn}.
Interestingly, all these deviations point to a coherent and consistent pattern, and can find a common explanation from new physics (NP) contributing to the Wilson coefficients $C_9$ (as was shown in e.g. Refs.~\cite{Descotes-Genon:2013wba,Altmannshofer:2013foa,Hurth:2013ssa,Hurth:2014vma}). 

While LFUV observables have theoretical uncertainties at the percent level (or below) due to the cancellation of hadronic uncertainties in the ratios, the rest of the $b \to s$ observables are subject to assumptions made for the non-local hadronic effects and generally suffer from larger theoretical uncertainties~\cite{Hurth:2016fbr}.
In this analysis, we consider the Minimal Supersymmetric Standard Model (MSSM)~\cite{WESS197439,FAYET1976159}, which predicts a superpartner particle (sparticle) to each SM field, together with an additional Higgs doublet. As supersymmetry (SUSY) is not observed at low energy scales, it needs to be a broken symmetry of nature. To preserve some of the nice features of supersymmetry, it should be ``softly'' broken, namely by introducing a SUSY-violating effective Lagrangian $\mathcal{L}_{SOFT}$, that contains all necessary couplings and masses, adding up to 105 new free parameters. \\
Until very recently, due to obvious computational challenges, the whole MSSM has been little studied. Indeed, more constrained SUSY models were devised to allow for doable calculations and computations, through well-motivated and seemingly reasonable, but not physically founded assumptions. Such models with simplifications at the GUT scale consider a handful of parameters, like the constrained MSSM (cMSSM)~\cite{Nilles:1983ge}. More recently, the phenomenological MSSM (pMSSM), which considers CP conservation and Minimal Flavour Violation (MFV) simplifications~\cite{MSSMWorkingGroup:1998fiq}, entered within computational reach with its 19 free parameters~\cite{Berger:2008cq,AbdusSalam:2009qd,Sekmen:2011cz,Arbey:2012ntc,Arbey:2011aa}. These models fail to provide a SUSY scenario fully compatible with the aforementioned flavour anomalies if R-parity is conserved~\cite{Mahmoudi:2014mja} (for R-parity violating models, see e.g. Refs.~\cite{Hu:2020yvs,BhupalDev:2021ipu}).

In this work, following our preliminary results~\cite{Boussejra:2021ych}, we will go one step further and consider for the first time a more general setup, based on the assumptions of the pMSSM but including in addition Non Minimal Flavour Violation (NMFV) in the squark sector, as a candidate for the explanation of the flavour anomalies in the $b \to s ll$ transitions. NMFV allows for sizeable Flavour Changing Neutral Currents (FCNC) effects coming directly from the squark mass matrices at the weak scale, whose off-diagonal entries are then considered as new free parameters with respect to MFV scenarios. 
We will first consider NMFV contributions to Wilson coefficients through the Mass Insertion Approximation (MIA) and show that the new FCNCs can highly affect the value of $C_9$ in some scenarios, while still being compatible with the rest of the $b \to s$ constraints. Then, we will present the first analytical calculation of the general contributions to $C_7$, $C_9$, $C_{10}$ and also $(g-2)_\mu$ in the full MSSM with 105 parameters, and their evaluation for particular NMFV scenarios with 42 parameters.

\newpage
The paper is organized as follows: Section~\ref{sec:theo} describes the theoretical context of our analysis.
In Section~\ref{sec:MIA} the flavour-violating parameters are introduced in the mass insertion approximation, and their new contributions to Wilson coefficients are defined. In Section~\ref{sec:num}, the numerical setup for our scans is presented. 
Section~\ref{sec:results_MIA} shows and discusses how the NMFV models may fit the flavour anomalies. In Section~\ref{sec:NMFV_MARTY}, we present the first full analytical evaluation of the Wilson coefficients and $(g-2)_\mu$, using \marty~\cite{Uhlrich:2020ltd,Uhlrich:2021ded}, in the MSSM and their evaluation in NMFV scenarios, confronting the results to the expected experimental values.
Finally, the conclusions are given in Section~\ref{sec:conclusion}.

\section{Theoretical Context}\label{sec:theo}
   
In the MSSM, the most general soft supersymmetry-breaking Lagrangian can be written as: 
$-\mathcal{L}_{SOFT} = -{\cal L}_{\rm gaugino} - {\cal L}_{\rm sfermions} -{\cal L}_{\rm Higgs}- {\cal L}_{\rm tril.}$, where the different terms are \cite{MSSMWorkingGroup:1998fiq,Martin:1997ns}
\begin{itemize}
  
    \item[$\bullet$] Mass terms for the gluinos, winos and binos:
     \begin{equation} \label{eq:Lsoft_gaugino}
    - {\cal L}_{\rm gaugino}=\frac{1}{2} \left[ M_1 \tilde{B}  
    \tilde{B}+M_2 \sum_{a=1}^3 \tilde{W}^a \tilde{W}_a +
    M_3 \sum_{a=1}^8 \tilde{G}^a \tilde{G}_a  \ + \ {\rm h.c.} 
    \right]\ ,
     \end{equation}
    where $\tilde{B}, \tilde{W}$ and $\tilde{G}$ are the bino, wino and gluino fields, respectively. 
    \item[$\bullet$] Mass terms for the scalar fermions: 
    \begin{equation} \label{eq:LSOFT_fermions}
    -{\mathcal L}_{\tilde{f}} = 
    \sum_{i,j=gen}  {\tilde{Q}}_i^{\dagger} (M^2_{\tilde {Q}})_{ij} {\tilde{Q}}_j+
    {\tilde{L}}_i^{\dagger} (M^2_{\tilde{L}})_{ij} {\tilde{L}}_j + 
    {\tilde{U}}^{\dagger}_{i} (M^2_{\tilde{U}})_{ij} {\tilde{U}}_{j} +
    {\tilde{D}}^{\dagger}_{i} (M^2_{\tilde{D}})_{ij} {\tilde{D}}_{j} +
    {\tilde{E}}^{\dagger}_{i} (M^2_{\tilde{E}})_{ij} {\tilde{E}}_{j}\ ,
    \end{equation}
    where $\tilde {Q}_i$ and $\tilde{L}_i$ are the left-handed squarks and sleptons, respectively, with their right-handed counterparts : $\tilde{U}, \tilde{D}$ and $\tilde{E}$ (no right-handed (s)neutrinos are assumed). The indices $i,j$ run over generation, and all scalar squared mass matrices are Hermitian.
    \item[$\bullet$] Mass and bilinear terms for the Higgs bosons: 
     \begin{equation} \label{eq:LSOFT_higgs}
    -{\cal L}_{\rm Higgs} = m^2_{H_u} H_u^{\dagger} H_u+m^2_{H_d}  H_d^{\dagger} 
    H_d +  \mu H_u.H_d + {\rm H.c.}\ ,
     \end{equation}
    where $\mu$ is the supersymmetric Higgs mass parameter.
    \item[$\bullet$] Trilinear couplings between sfermions and Higgs bosons 
     \begin{equation}
    -{\cal L}_{\rm tril.}= 
    {\sum_{i,j=gen}  A^u_{ij} Y^u_{ij}  {\tilde{u}}_{R_i} H_u. 
    {\tilde{Q}}_j+
    A^d_{ij} Y^d_{ij}  {\tilde{d}}_{R_i} H_d.{\tilde{Q}}_j
    +A^l_{ij} Y^l_{ij} {\tilde{l}}_{R_i} H_u.{\tilde{L}}_j +\  {\rm H.c.} 
    \ , }
     \label{eq:LSOFt:tril}
     \end{equation}
    where $A^f_{ij}$ are the general $3 \times 3$ complex soft SUSY-breaking scalar trilinear coupling matrices between Higgs fields ($H_u$,$H_d$) and sfermions, in generation basis.
\end{itemize}

Several mixing effects arise in the general MSSM. In particular, the electroweak gauginos mix together with the higgsinos and give rise to the chargino and neutralino mass eigenstates. The chargino mixing matrix in the weak eigenstate $(\tilde{W}^+,\tilde{H}_u^+,\tilde{W}^-,\tilde{H}^-_d)$ basis is given by
\begin{equation}
M_\chi =\left(
\begin{array}{cc}
M_2 & \sqrt{2}M_W \sin \beta \\
 \sqrt{2}M_W \cos \beta &\mu
\end{array}
\right)\ ,
\label{eq:charg_matrix}
\end{equation}
where  $\mu$ is the Higgs quadratic coupling and $M_2$ the soft SUSY-breaking wino mass. The $\beta$ parameter is related to the vacuum expectation values of the two Higgs doublets present in the MSSM by 
\begin{equation}
    \tan \beta \equiv \frac{v_u}{v_d} \ ,
\end{equation}
with $v_u =  \expval{H^0_u} = v \sin \beta $ and $v_d = \expval{H^0_d} = v \cos \beta $.\\

The $2 \times 2$ unitary matrices $U$ and $V$ which diagonalize the chargino mass matrix $M_{\chi}$ are defined as
\begin{equation}
U^* M_{\chi} V^{-1} ={\rm diag}(M_{\chi^\pm_1},M_{\chi^\pm_2}) \ .
\end{equation}
Their explicit expressions can be found in e.g. Refs.~\cite{drees2005theory,Lunghi:1999uk,MSSMWorkingGroup:1998fiq}. \\

In the other sectors, the MFV hypothesis limits the mixing of squarks to the third generation only. This approach is still widely used in the study of the MSSM.
If the MFV hypothesis is relaxed for other generations, a rich mixing dynamic arises. 
Concentrating on the squark sector in such a NMFV model, and starting from the Lagrangian in~\eqref{eq:LSOFT_fermions}, one can define the super-CKM (sCKM) basis so that it  rotates the (s)quarks' superfields in flavour space, making the quark mass matrices $m_{u,d}$ diagonal. 
This flavour alignment between quarks and squarks does not imply diagonal squark mass matrices, and can yield substantial flavour-changing effects.\\
In the same manner as Ref.~\cite{drees2005theory}, let
\begin{equation}
    \boldsymbol{\tilde{f}} \equiv 
        \begin{pmatrix}
            \tilde{f_L} \\
            \tilde{f_R} \\
        \end{pmatrix}\ ,
\end{equation}\label{eq:f_vector}%
be a six-component vector, where $\tilde{f_L},\tilde{f_R}$ are spanning generation space.
We can therefore write the $6 \times 6 $ flavour mixed squared fermion mass matrices as 
\begin{equation}
    \mathcal{M}^2_{\boldsymbol{\tilde{f}}} = 
    \begin{pmatrix}
        \mathcal{M}^2_{\boldsymbol{\tilde{f}}_{LL}} & \mathcal{M}^2_{\boldsymbol{\tilde{f}}_{RL}} \\
        \mathcal{M}^2_{\boldsymbol{\tilde{f}}_{LR}} & \mathcal{M}^2_{\boldsymbol{\tilde{f}}_{RR}} \\
    \end{pmatrix} \ .
    \label{eq:fermions_mass_matrices}
\end{equation}
Collecting all sfermion mass terms in~\eqref{eq:LSOFT_fermions} and using~\eqref{eq:fermions_mass_matrices}
\begin{equation}
    - \mathcal{L}_{\mathcal{M}^2_{\boldsymbol{\tilde{f}}}} = \sum_{\boldsymbol{\tilde{f}}} \boldsymbol{\tilde{f}}^{\dagger} \mathcal{M}^2_{\boldsymbol{\tilde{f}}} \boldsymbol{\tilde{f}}\ .
\end{equation}
This defines the relevant mass matrices for the squark sector: $\mathcal{M}^2_{\boldsymbol{\tilde{u}}}, \mathcal{M}^2_{\boldsymbol{\tilde{d}}}$, in the corresponding bases $(\tilde u_L,\tilde c_L,\tilde t_L,\tilde u_R,\tilde c_R,\tilde t_R)$ and
$(\tilde d_L,\tilde s_L,\tilde b_L,\tilde d_R,\tilde s_R,\tilde b_R)$ . Their complete expressions can be found e.g. in Ref.~\cite{Allanach:2008qq}, and a thorough analysis of the various terms at play can be found in Ref.~\cite{drees2005theory}.
Following Ref.~\cite{Allanach:2008qq}, we define 
\begin{equation}
\label{eq:squark_mass_matrices}
\begin{split}
     \mathcal{M}^2_{\boldsymbol{\tilde{d}}} ~&=~
    \begin{pmatrix}
        M^2_{\tilde{Q}} + m^2_{d} + D_{\tilde{d},L}
        ~~&~~
        \frac{v_d}{\sqrt{2}} T_d^\dag - m_d \mu \tan\beta \\
        \frac{v_d}{\sqrt{2}} T_d - m_d \mu^* \tan\beta  
        ~~&~~
        M^2_{\tilde{D}} + m^2_{d} + D_{\tilde{d},R}
    \end{pmatrix} \, , \\ 
    \mathcal{M}^2_{\boldsymbol{\tilde{u}}} ~&=~ 
    \begin{pmatrix}
        V_{\rm CKM} M^2_{\tilde{Q}} V^{\dag}_{\rm CKM} + m^2_{u} + D_{\tilde{u},L}
        ~~&~~
     \frac{v_u}{\sqrt{2}} T_u^\dag - m_u \frac{\mu}{\tan\beta} \\
     \frac{v_u}{\sqrt{2}} T_u - m_u \frac{\mu^*}{\tan\beta}
     ~~&~~
     M^2_{\tilde{U}} + m^2_{u} + D_{\tilde{u},R}
    \end{pmatrix} \ ,
    \end{split}
\end{equation}
where the $M^2_{\tilde{U}}, M^2_{\tilde{D}}$ and $M^2_{\tilde{Q}}$ are the soft breaking squark masses defined in~\eqref{eq:LSOFT_fermions}, and $m_{u,d}$ are the diagonal up- and down-type quark masses. The various $D$ terms are given by 
\begin{equation}
    {D_f}_{LL,RR} = \cos 2\beta \; m_Z^2 \, (T_f^3 - Q_f \sin^2 \theta_W ) \mathbb{1}_3 \ ,
\end{equation}
which are obviously flavour diagonal. \\
Finally, the $T_{u,d}$ terms are related to the trilinear quark-squark-Higgs couplings in~\eqref{eq:LSOFt:tril} by
\begin{align}
    (T_u)_{ij} ~&~\equiv  ( A^u Y^u)_{ij} \ , \\ 
    (T_d)_{ij} ~&~\equiv  ( A^d Y^d)_{ij} \ .
\end{align}

The final mass-ordered squark mass eigenstates are obtained by introducing the unitary transformation to the matrices in~\eqref{eq:squark_mass_matrices}: 
\be
  {\rm diag}\big(m^2_{\tilde{q}_1},m^2_{\tilde{q}_2},\dots,m^2_{\tilde{q}_6}\big)
     ~=~ {\cal R}_{\tilde{q}} {\cal M}_{\tilde{q}}^2 {\cal R}^{\dag}_{\tilde{q}}\ ,
  ~~\text{for}~~  q=u,d, ~~\text{and} ~~ m^2_{\tilde{q}_1} < \dots < m^2_{\tilde{q}_6}\ ,
\ee
with the matrices ${\cal R}_{\tilde{u},\tilde{d}}$ containing the flavour decomposition information of the mass-ordered  squark mass eigenstates: 
\be\bsp
  \bpm \tilde u_1&  \tilde u_2&  \tilde u_3&  \tilde u_4&  \tilde u_5 &  \tilde
    u_6 \epm^t =&\ {\cal R}_{\tilde u} \bpm \tilde u_L&  \tilde c_L&  \tilde t_L&  \tilde u_R & 
    \tilde c_R&  \tilde t_R \epm^t \ ,\\
  \bpm \tilde d_1&  \tilde d_2&  \tilde d_3&  \tilde d_4&  \tilde d_5 &  \tilde
    d_6 \epm^t =&\ {\cal R}_{\tilde d} \bpm \tilde d_L&  \tilde s_L&  \tilde b_L&  \tilde d_R & 
    \tilde s_R&  \tilde b_R \epm^t \ .
\esp\ee \\

Transformations between mass and flavour eigenstates are needed to perform phenomenological analyses on the model, as its parameters cannot be accessed directly from the mixed final eigenstates. The complexity of such analyses grows rapidly with the allowed mixings and free parameters. 
The computational challenge is such that a complete analysis of the most general MSSM with its 105 free parameters is not feasible. 
We propose two approaches: one within the so-called Mass Insertion Approximation with 28 free parameters, and then within a subset of the MSSM including NMFV with 42 parameters.

\section{The Mass Insertion approach to the NMFV-MSSM}\label{sec:MIA}
  
    \begin{figure}[ht]
        \begin{center}
            \includegraphics[width=0.3\linewidth]{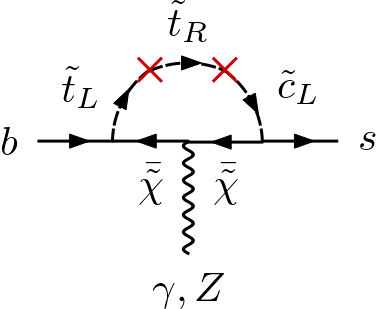}
            \hspace{1cm}
            \includegraphics[width=0.3\linewidth]{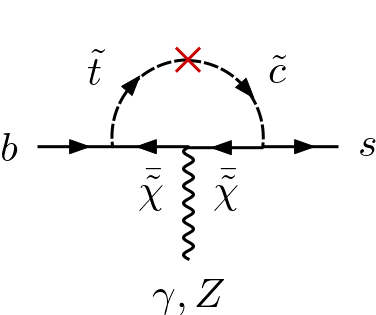}\\
            \vspace{1cm}
            \includegraphics[width=0.3\linewidth]{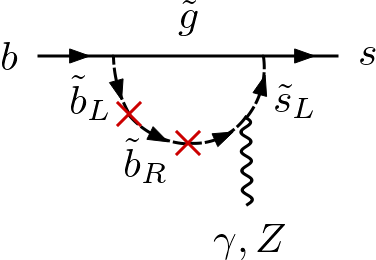}
            \hspace{1cm} \vspace{2cm}
            \includegraphics[width=0.3\linewidth]{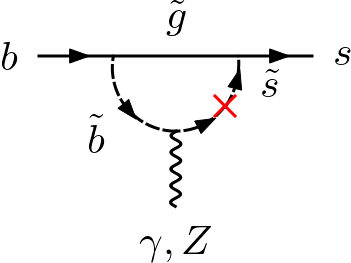}
            \caption[]{Some of the relevant penguin diagrams for $b\rightarrow
            s \ell^+ \ell^-$. The red cross
            indicates a Mass Insertion. First row diagrams are based on chargino interactions.
            The ones at the bottom consider gluino interactions.}\label{fig:sfigs}
        \end{center}
    \end{figure}
    
    \begin{figure}[ht]
        \begin{center}
            \includegraphics[scale=0.3]{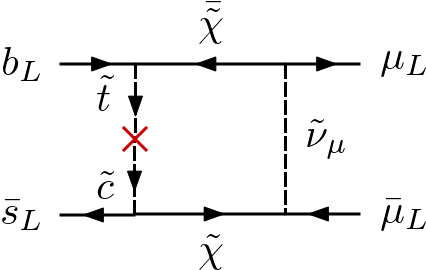}
            \caption[]{Relevant box diagram for $b\rightarrow s \ell^+
            \ell^-$. The red cross
            indicates a Mass Insertion. }
            \protect\label{fig:box}
        \end{center}
    \end{figure}
The usual approach when studying NMFV effects in the MSSM is to use the MIA approach, introduced as early as 1989 in Ref.~\cite{Gabbiani:1988rb}. \\
The MIA originates as a diagrammatic technique~\cite{Gabbiani:1988rb,Gabrielli:1995bd}, allowing us to choose a basis where the quark-squark-neutral gaugino couplings are flavour diagonal. The flavour-changing effects are provided by non-diagonal contributions in the sfermion propagators, as shown in Figs.~\ref{fig:sfigs} and~\ref{fig:box}. The new SUSY contributions (to e.g. Wilson coefficients) are then proportional to the various off-diagonal elements. 

The MIA is also defined algebraically through the Flavour Expansion Theorem (FET)~\cite{Dedes:2015twa} the main features of which we will summarize in the following.
All sfermion squared mass matrices $M$ can be decomposed as a sum of a diagonal ${\rm diag}(M_{ii}) \equiv M^d_i$ and a non-diagonal $\hat{M}_{ij}$ matrix. Calculating loop amplitudes requires evaluating Hermitian matrix functions $f(M)$ of the involved mass matrices, which can be expanded following the FET's conditions:
\begin{equation}
    f(M)_{ij} = \delta_{ij} f(M^d_i) + f^{[1]}(M^d_i,M^d_j) \hat{M}_{ij} + 
    \sum\limits_{n_1} f^{[2]}(M^d_i,M^d_j,M^d_{n_1}) \hat{M}_{i n_1}\hat{M}_{j n_1} + \dots \ , 
\end{equation}
where the divided difference $f^{[k]}$ functions are defined in Ref.~\cite{Dedes:2015twa}. 

This expansion expresses the loop quantities such as Wilson coefficients in terms of the flavour-violating off-diagonal entries in the squark squared mass matrices. The following dimensionless ratio is usually introduced to define the mass insertions: 
\begin{equation}
    \delta^{\tilde{f}}_{ij} = \frac{(M^2_{\tilde{f}})_{ij}}{\sqrt{(M^2_{\tilde{f}})_{ii}(M^2_{\tilde{f}})_{jj}}}
\end{equation}
where $M^2_{\tilde{f}}$ is one of the fermion soft-breaking matrices in~\eqref{eq:LSOFT_fermions}. 
As the full sfermion mass matrix is actually a $6 \times 6$ matrix spanning both generation and chirality indices~\eqref{eq:fermions_mass_matrices}, the actual mass insertion parameter is of the form $(\delta^{\tilde{f}}_{ij})_{AB}$ where $i,j$ are generation indices, and $(AB) \in \{LL,LR,RL,RR\}$.

In this framework, we define the relevant Mass Insertions (MI) for our study. To be consistent with the constraints from Kaon observables~\cite{Ciuchini:2007ha,Gabrielli:1995bd,drees2005theory}, every off-diagonal element involving a first-generation squark is neglected. The relevant $(\delta^{\tilde{f}}_{23})_{AB}$ are
\be\bsp
  \delta^d_{LL}=
     \frac{(M^2_{\tilde{Q}})_{23}}{(M_{\tilde{Q}})_{22}( M_{\tilde{Q}})_{33}}\ ,
  \quad
   \delta^u_{RR}=
     \frac{(M^2_{\tilde{U}})_{23}}{(M_{\tilde{U}})_{22}( M_{\tilde{U}})_{33}}\ ,
  \quad
  \delta^d_{RR}=
     \frac{(M^2_{\tilde{D}})_{23}}{(M_{\tilde{D}})_{22}( M_{\tilde{D}})_{33}}\ , \\
  \delta^u_{RL}= \frac{v_u}{\sqrt{2}}
     \frac{(T_{u})_{23}}{(M_{\tilde{Q}})_{22}( M_{\tilde{U}})_{33}}\ ,
  \qquad
  \delta^u_{LR}= \frac{v_u}{\sqrt{2}}
     \frac{(T_{u})_{32}}{(M_{\tilde{Q}})_{33}( M_{\tilde{U}})_{22}}\ ,
  \qquad\qquad \\
  \delta^d_{RL}= \frac{v_d}{\sqrt{2}}
     \frac{(T_{d})_{23}}{(M_{\tilde{Q}})_{22}( M_{\tilde{D}})_{33}}\ ,
  \qquad
  \delta^d_{LR}= \frac{v_d}{\sqrt{2}}
     \frac{(T_{d})_{32}}{(M_{\tilde{Q}})_{33}( M_{\tilde{D}})_{22}}\ .
  \qquad\qquad
\esp\label{eq:NMFVparams}\ee
For the $\delta^u_{LL}$ insertion, following the definition of $ \mathcal{M}^2_{\boldsymbol{\tilde{u}}}$ in~\eqref{eq:squark_mass_matrices}, we express it in terms of the soft-breaking squark mass matrix $M^2_{\tilde{Q}}$ as
\be
\delta^u_{LL}=
     \frac{(V_{\rm CKM} M^2_{\tilde{Q}} V^{\dag}_{\rm CKM})_{23}}{(V_{\rm CKM} M_{\tilde{Q}} V^{\dag}_{\rm CKM})_{22}( V_{\rm CKM} M_{\tilde{Q}} V^{\dag}_{\rm CKM})_{33}}
     \label{eq:delta_u_LL}\ .
\ee
All the relevant NMFV contributions to the $C_7,C_9$ and $C_{10}$ Wilson coefficients are given in Appendix~\ref{sec:wilson_coeff}.

\section{Numerical Setup}\label{sec:num}
In what follows, we present a study of NMFV contributions to the $b \rightarrow s ll$ processes in terms of Wilson coefficients (given in~\ref{sec:wilson_coeff}) and mass insertions. The model used is an extension of the phenomenological MSSM (pMSSM), where the new contributions arise from additional flavour violation sources in the form of mass insertions.
No new sources of CP violation in ${\cal L_{\rm SOFT}}$ with respect to the pMSSM are included, and the degeneracy between the first and second generations of squarks is kept.\\
The third-generation trilinear interactions $A_t, A_b$ and $A_{\tau}$ are allowed to vary, while the others are set to zero.
As we will see in Section~\ref{sec:SLHA2_combined_analysis}, the slepton sector contribution to $(g-2)_{\mu}$ can be completely decoupled from the squark sector analysis. Therefore, no flavour-violating effect is turned on in the slepton sector, as they do not contribute to the $b \rightarrow s l l$ observables.
The Standard Model sector parameters are given in Table~\ref{tab:cost}.

\begin{table}[!h]
\begin{center}     
\begin{tabular}{@{}ll@{}}
    \toprule
    SM parameter    &   Value \\ \midrule
    $m_t$ & 173.8 GeV \\
    $m_b$ & 4.8 GeV \\
    $m_c$ & 1.4 GeV \\
    $m_s$ & 125 MeV \\
    $M_B$ & 5.27 GeV\\
    $\a_s(m_Z)$ & 0.119 \\
    $1/\a_{el}(m_Z)$ &128.9 \\
    $\sin^2 \theta_W$ & 0.2334 \\  \bottomrule 
\end{tabular}
\caption{SM parameters' values used in this study.}
\label{tab:cost}
\end{center}
\end{table}

The 28 input parameters for our model and their ranges (pMSSM+MIA) are given in Tables~\ref{tab:param1} and~\ref{tab:param_deltas}. They are randomly sampled from a uniform distribution. The spectrum is calculated at the electroweak scale using the code \softsusy ~\cite{Allanach:2001kg}. Following Ref.~\cite{Lunghi:1999uk}, we have introduced an average squark mass (over the first two generations), obtained from the resulting spectra, and used it to compute the various Wilson coefficients: 
\begin{equation}
     M_{sq} \equiv \frac{1}{8} \sum_{\rm{squarks}} m_{\rm{squarks}}\ .
\end{equation}
We then use \superiso~\cite{Mahmoudi:2007vz,Mahmoudi:2008tp,Mahmoudi:2009zz,Neshatpour:2021nbn} to compute all relevant pMSSM contributions from the obtained spectrum. The additional NMFV contributions to Wilson coefficients are computed following the formulae presented in the Appendix~\ref{sec:wilson_coeff}. \\

\begin{table}[hbt!] 
    \begin{minipage}{0.45\linewidth}\centering
     \begin{tabular}{@{}ll@{}}
        \toprule
            Parameter  & Range            \\ \midrule
            $M_1$       & $[50,5000]$        \\
            $M_2$       & $[50,5000]$        \\
            $M_3$       &  $[50,5000]$       \\
            $m_A$        & $[50,5000]$       \\
            $\tan\beta$ & $[2,60]$           \\
            $\mu$       & $[-10^4,10^4]$ \\
            $A_t, A_b, A_{\tau}$       & $[-10^4,10^4]$ \\
            $M_{\tilde{q}_{1L}},M_{\tilde{q}_{3L}}$& [50,5000] \\
            $M_{\tilde{u}_R}, M_{\tilde{d}_R}, M_{\tilde{t}_R}, M_{\tilde{b}_R} 
            $    & [50,5000] \\
            $M_{\tilde{e}_L},M_{\tilde{\tau}_L},M_{\tilde{e}_R},M_{\tilde{\tau}_R}$ & [50, 5000] \\ \bottomrule
        \end{tabular}%
        \caption{Allowed ranges for the 19 pMSSM soft-breaking parameters.} \label{tab:param1}
        \end{minipage}\qquad
    \begin{minipage}{0.45\linewidth}\centering
      \begin{tabular}{@{}ll@{}}
                \toprule
                Parameter          & Range     \\ \midrule
                $(\delta^{\tilde{u}}_{23})_{LR}$  & [-1,1] \\
                $(\delta^{\tilde{u}}_{23})_{LL}$  & [-1,1] \\
                $(\delta^{\tilde{u}}_{33})_{LR}$  & [-1,1] \\
                $(\delta^{\tilde{d}}_{23})_{LL}$  & [-1,1] \\
                $(\delta^{\tilde{d}}_{23})_{RR}$  & [-1,1] \\
                $(\delta^{\tilde{d}}_{23})_{RL}$  & [-1,1] \\
                $(\delta^{\tilde{d}}_{23})_{LR}$  & [-1,1] \\
                $(\delta^{\tilde{d}}_{33})_{RL}$  & [-1,1] \\
                $(\delta^{\tilde{d}}_{33})_{LR}$  & [-1,1] \\ \bottomrule
        \end{tabular}
        \caption{Additional NMFV input parameters in the MIA.}
        \label{tab:param_deltas}
    \end{minipage}
\end{table}

\subsection{Constraints}\label{sec:constr}
In the following, we give the constraints considered in our study, both during and after the sampling of the parameter space.

First, no tachyonic spectra are kept. This is a built-in condition in many spectrum calculators such as \softsusy\, which is enforced during execution. \\
We discard any spectra with a charged Lightest Supersymmetric Particle (LSP) to ensure the possibility for the LSP (often the lightest neutralino) to be a viable dark matter candidate.
We impose further the latest available mass limits from supersymmetric searches given by Ref.~\cite{ParticleDataGroup:2020ssz}.
No additional {\it ab} initio constraints are imposed, in order to keep the study as general as it can be. \\

As the SLHA1~\cite{Skands:2003cj} file format does not implement flavour mixing, the spectrum yielded by \softsusy\ is obtained without considering the flavour-violating MIA parameters. Therefore, the spectrum considered here is pMSSM-like. The $\delta$s are considered as additional free parameters, that do not intervene in the computation of the spectrum. This can be justified {\it a} posteriori by considering constraints on the MI, as the approximation should be valid if they are small enough with respect to the diagonal mass parameters. 

The following limits on the MIA parameters are also considered a posteriori:
\begin{itemize}
    \item To avoid tachyonic sparticles, all the MI parameters' ranges are reduced to 
        \begin{equation}\label{eq:delta_cstr_tachyons}
            \abs{\delta^{\tilde{f}}_{AB}} < 0.85 \ .
        \end{equation}
    \item From vacuum stability arguments~\cite{Casas:1996de, Lunghi:1999uk}
    \begin{equation}\label{delta_cstr_vacCCB}
         \abs{(\delta^{u}_{23})_{LR}} < m_t \frac{\sqrt{2 M_{sq}^2 + 2 <m_{\tilde{l}}^2 >}}{M_{sq}^2} \simeq \frac{m_t}{M_{sq}} \ .
    \end{equation}
\end{itemize}
The flavour-violating parameters that contribute the most to $C_9$ are $(\delta^{u}_{23})_{LL}$ and $(\delta^{u}_{23})_{LR}$, in the chargino penguin diagrams such as the ones shown in Fig.~\ref{fig:sfigs}, which are mainly constrained by~\eqref{eq:delta_cstr_tachyons} and~\eqref{delta_cstr_vacCCB}. On the other hand, in the $\tilde{d}$ sector, the gluino loops contribute mostly to $C_7$ which is already strongly limited by experimental data. Therefore, considering all double mass insertions as negligible, no constraints on the other MI parameters are imposed. A comprehensive discussion of the allowed ranges for these parameters can be found in Ref.~\cite{DeCausmaecker:2015yca}.\\
Finally, all spectra should be considered with particular care, as flavour mixing can significantly affect the squark masses and their expected signal topologies at colliders. Also, the recast of LHC limits for general MSSM models is a non-trivial task~\cite{Bernigaud:2018vmh,Alguero:2021dig}, which goes beyond the scope of this study. Therefore, no particular limits on the sparticle masses are considered, apart from the model-independent ones present in Ref.~\cite{ParticleDataGroup:2020ssz}. \\ 
    
\section{Results and Discussion}\label{sec:results_MIA}
\begin{figure}[t!]
    \begin{subfigure}{0.5\linewidth}
    \includegraphics[width=0.87\linewidth]{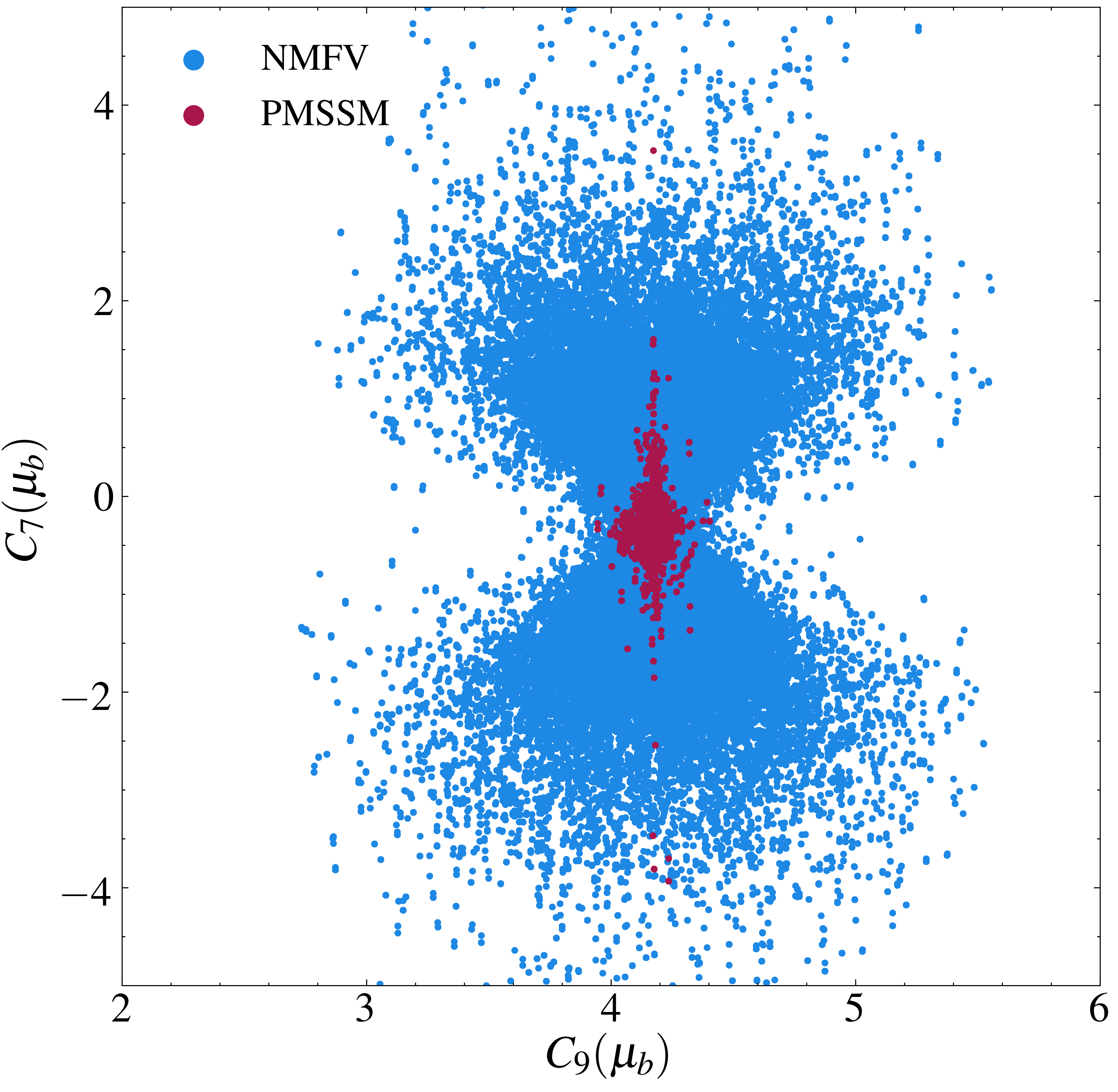}
     \end{subfigure}
     \begin{subfigure}{0.5\linewidth}
     \includegraphics[width=0.9\linewidth]{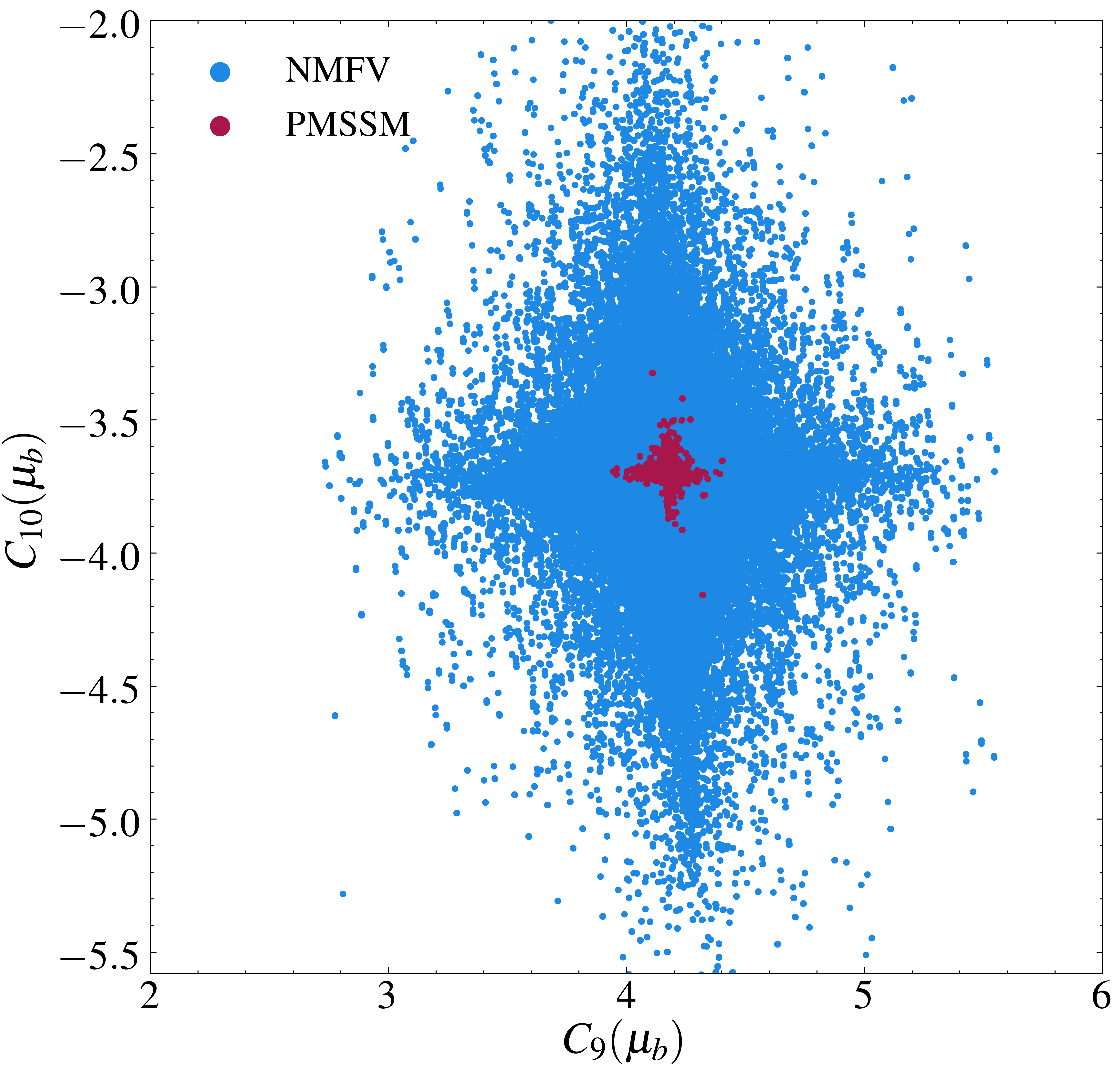}
     \end{subfigure}
    \caption{Combined distributions of the scanned points in the $(C_9,C_7)$ and $(C_9,C_{10})$ planes. The blue distribution is calculated in the NMFV augmented pMSSM, and the corresponding pMSSM points are shown in red.}
    \label{fig:C9C7C10 scatter}
\end{figure}

The mass insertions allow new sources of FCNC, which give sizeable contributions to flavour observables by significantly shifting the relevant Wilson coefficients. In Fig.~\ref{fig:C9C7C10 scatter}, we present the scan results with about 2 million model points. We can see an oyster-shaped spread of the pMSSM distribution upon turning on the NMFV contributions in the $(C_9,C_7)$ plane. In the $(C_9, C_{10})$ case, we can see an isotropic spread of the pMSSM distribution in all quadrants, indicating a homogeneous behaviour of the two Wilson coefficients under flavour violation in the squark sector. On the other hand, in the $(C_9, C_7)$ case, the largest contribution to $C_9$ can be obtained by shifting $C_7$ significantly from its SM value, which is strongly constrained by the $b \rightarrow s \gamma$ data. However, it is clear from the impressive spread that the flavour anomalies can be given a satisfying answer using this framework, while still having reasonable values for $C_7$. \\

\begin{figure}[h!]
    \begin{subfigure}{0.5\linewidth}
    \includegraphics[width=0.95\linewidth]{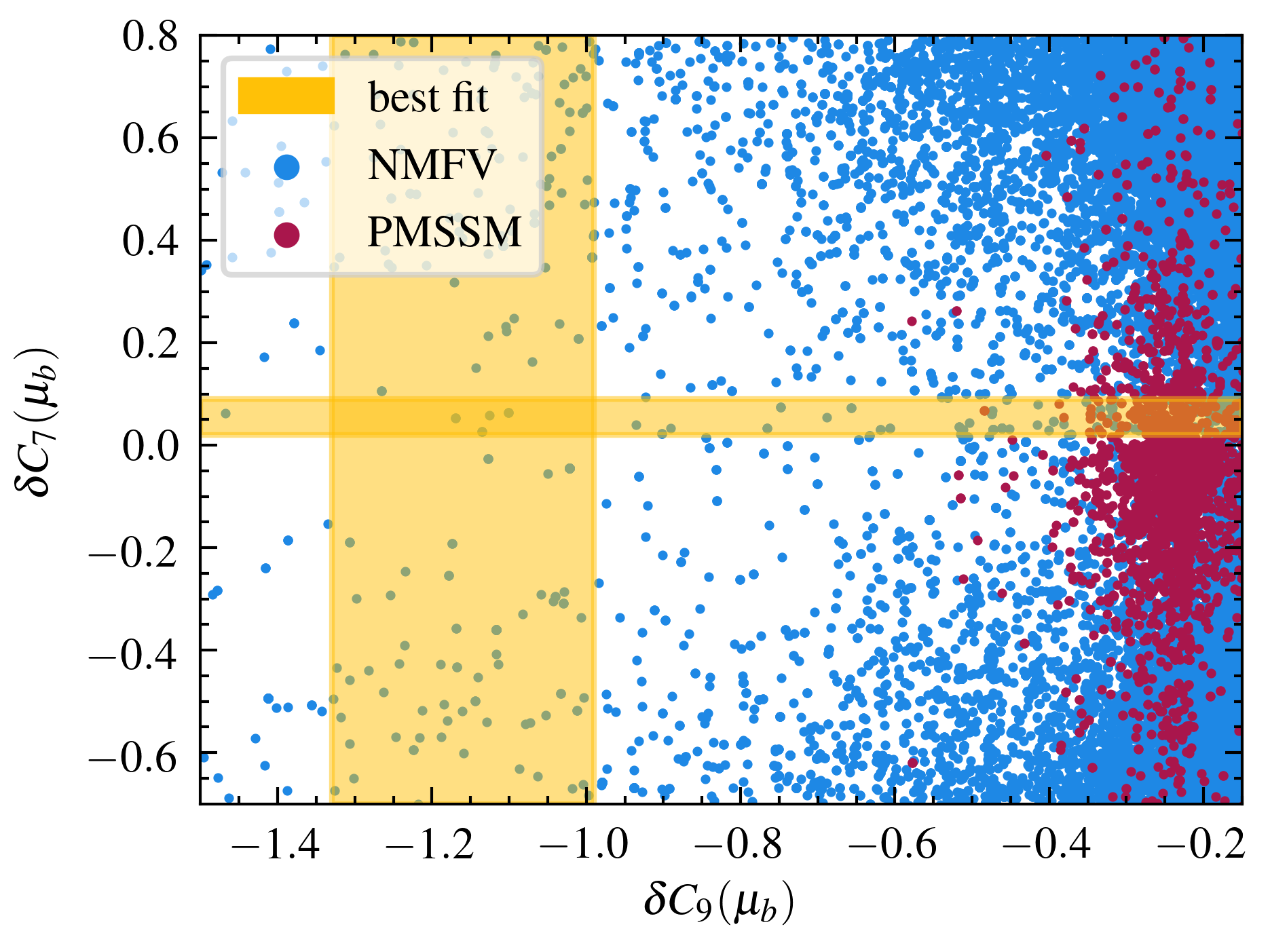}
    \caption{}
    \label{fig:C9C7 scatter best fit}
    \end{subfigure}
    \begin{subfigure}{0.5\linewidth}
    \includegraphics[width=0.95\linewidth]{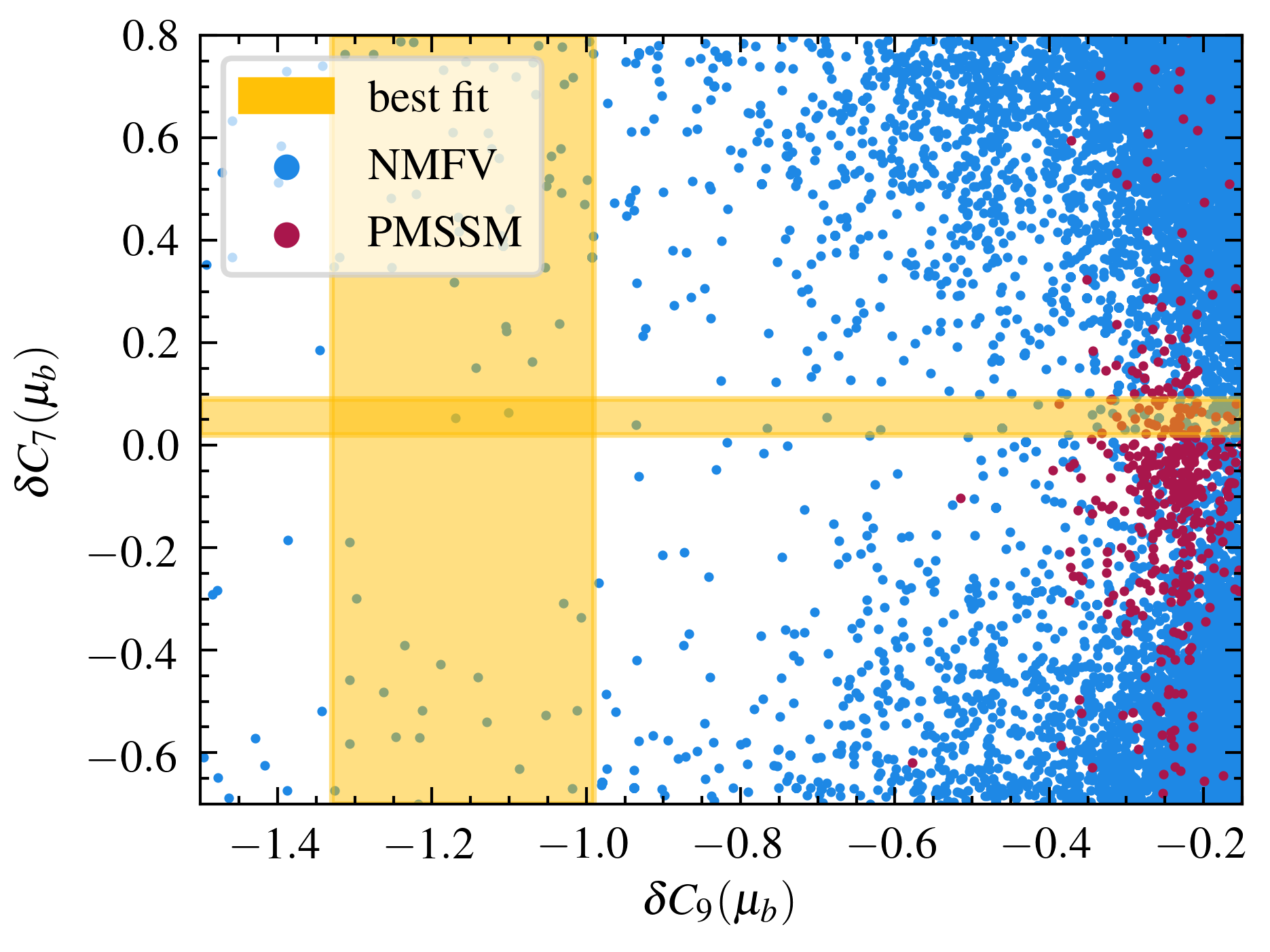}
    \caption{}
    \label{fig:C9C7_bf_cuts}
    \end{subfigure}
    \caption{Combined distribution of the scanned model points in the $(\delta C_9, \delta C_7)$ plane for the whole sample (a) and after applying cuts (b), where $\delta C_i(\mu_b) = C_i^{\textrm{NMFV,pMSSM}}(\mu_b) - C_i^{\textrm{SM}}(\mu_b)$. The orange bands represent the $1\sigma$ best-fit regions from Ref.~\cite{Hurth:2021nsi}.}
\end{figure}

\begin{figure}
    \centering
    \includegraphics[width=0.6\linewidth]{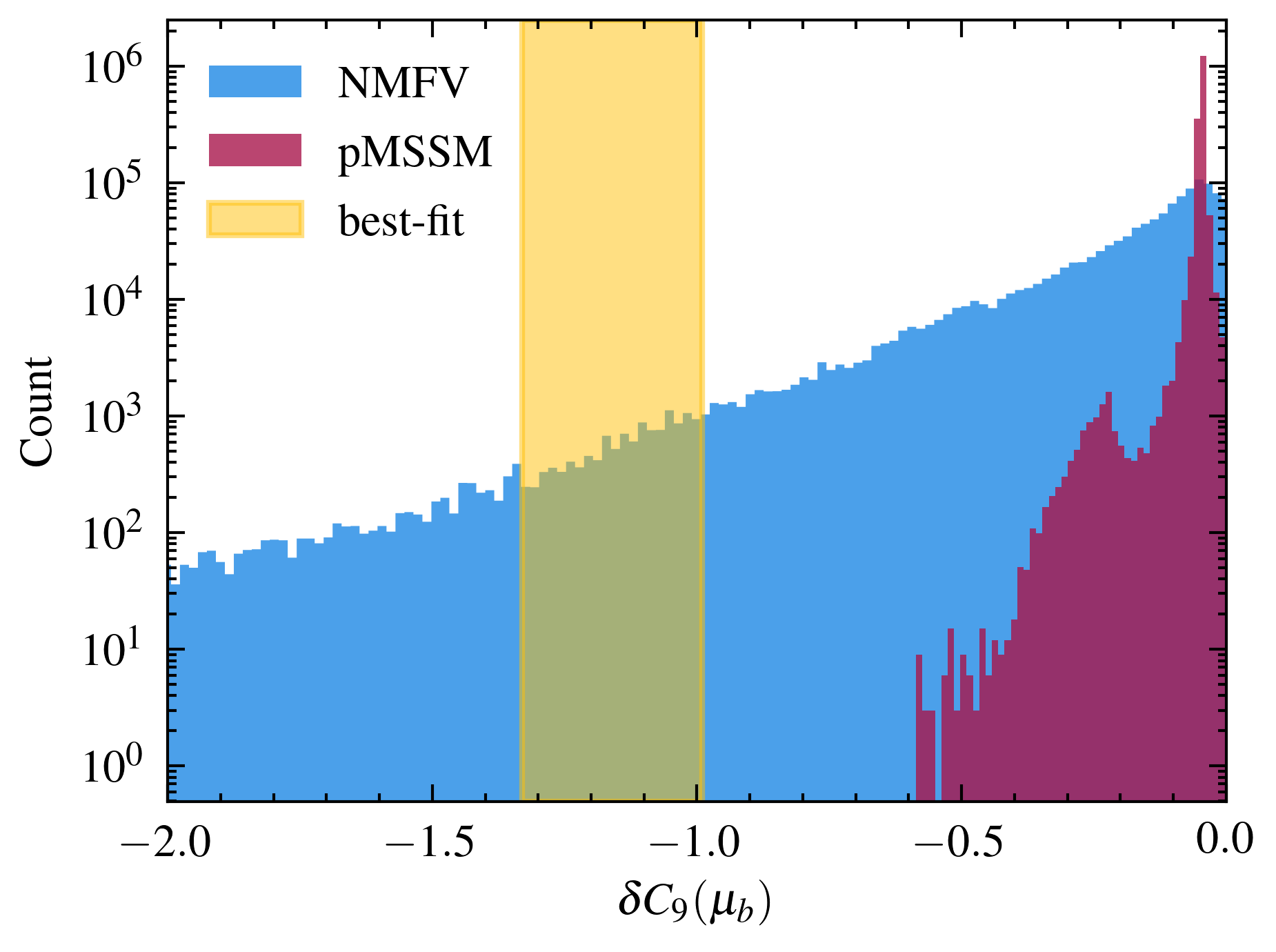}
    \caption{Compared distribution of $\delta C_9(\mu_b) \leq 0$ for both the pMSSM and our NMFV model. The SM value is shown at the dashed black line, and the best-fit patches are shown in orange. The bins of each histogram differ to show the features of each model.}
    \label{fig:hist_C9}
\end{figure}

In Fig.~\ref{fig:C9C7 scatter best fit}, a zoom in the region of interest in the $(\delta C_9, \delta C_7)$ plane is presented, together with the global best-fit patches from Ref.~\cite{Hurth:2021nsi}. $\delta C_i$ is defined as $C_i^{\textrm{NMFV}} - C_i^{\textrm{SM}}$. The pMSSM distribution is shown in red, and the corresponding NMFV points are shown in blue. Imposing the constraints discussed in Section~\ref{sec:constr} yields Fig.~\ref{fig:C9C7_bf_cuts} with 1 721 remaining points. We can see that even if the highest density of model points can be found away from the $C_7$ best-fit region, the presented NMFV model succeeds in proposing valid scenarios. In particular, several points seem to completely account for the flavour anomalies in the $B$ sector, but further exploration of the full model spectrum is necessary.
Also, it is clear that the pMSSM alone cannot give sufficient contributions to $C_9$ and $C_7$: the red distribution can at most account for half of the required shift in $C_9$ to explain the anomalies, with no other constraints imposed on the pMSSM parameters. Indeed, Fig.~\ref{fig:hist_C9} clearly shows this feature, where we can see the spread of $C_9$ for both pMSSM and NMFV models, with no particular constraints on the sampled points. The pMSSM, while being able to provide compelling shifts, fails to fully account for the anomalies (best fit given in~Refs.\cite{Hurth:2017hxg,Arbey:2019duh,Hurth:2021nsi}) as was shown already in~Ref.\cite{Mahmoudi:2014mja}, whereas the NMFV is capable of providing hundreds of compatible scenarios if no other constraints are considered.\\ 

\begin{figure}[h]
    \begin{subfigure}{0.5\textwidth}
      \includegraphics[width=\linewidth]{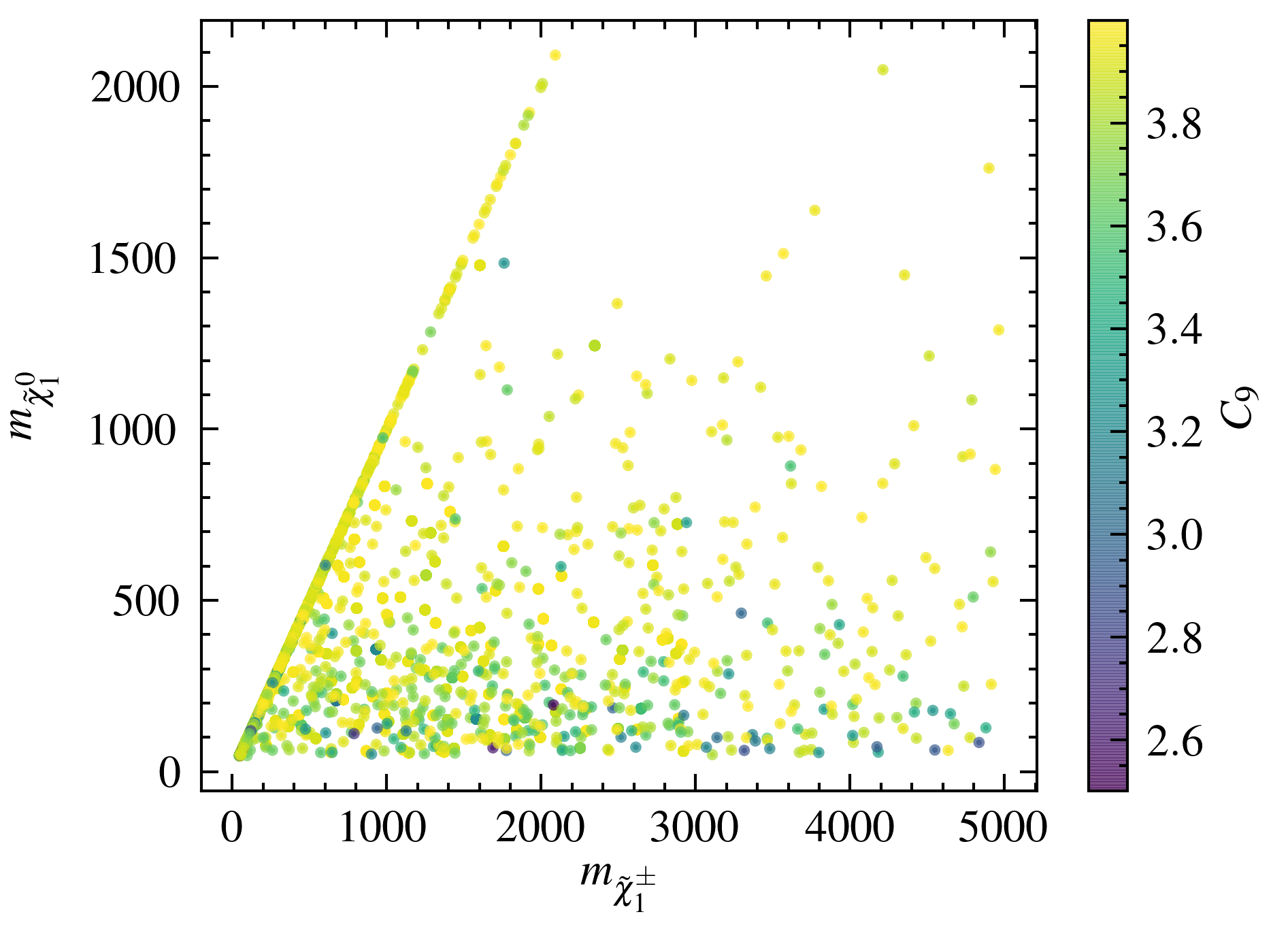}
      \caption{}
      \label{sfig:ch_neut_uncut}
    \end{subfigure}
    \begin{subfigure}{0.5\textwidth}
        \includegraphics[width=\linewidth]{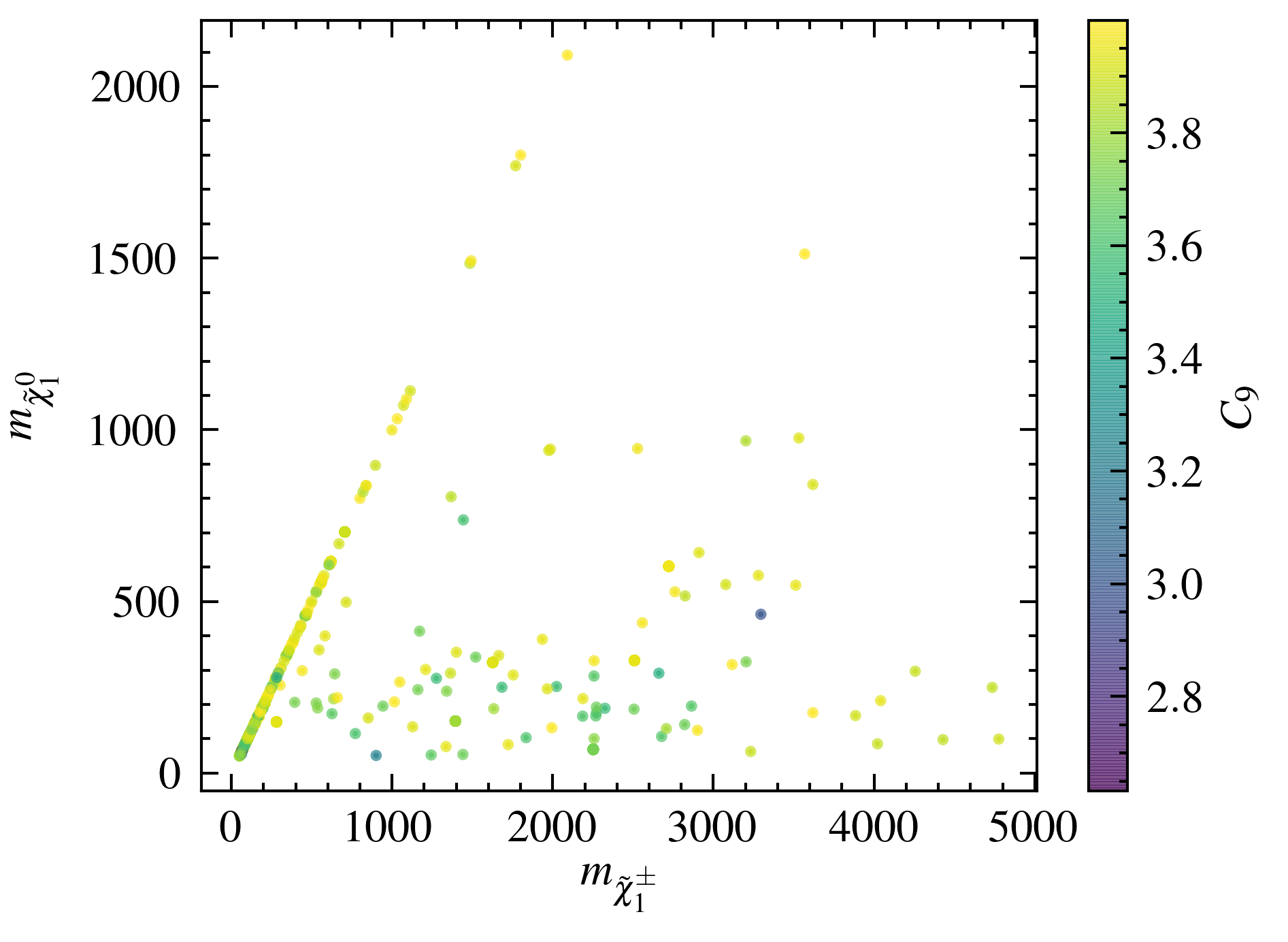}
        \caption{}
        \label{sfig:ch_neut_cut}
    \end{subfigure}
    \newline
    
    \begin{subfigure}{0.5\textwidth}
        \includegraphics[width=\linewidth]{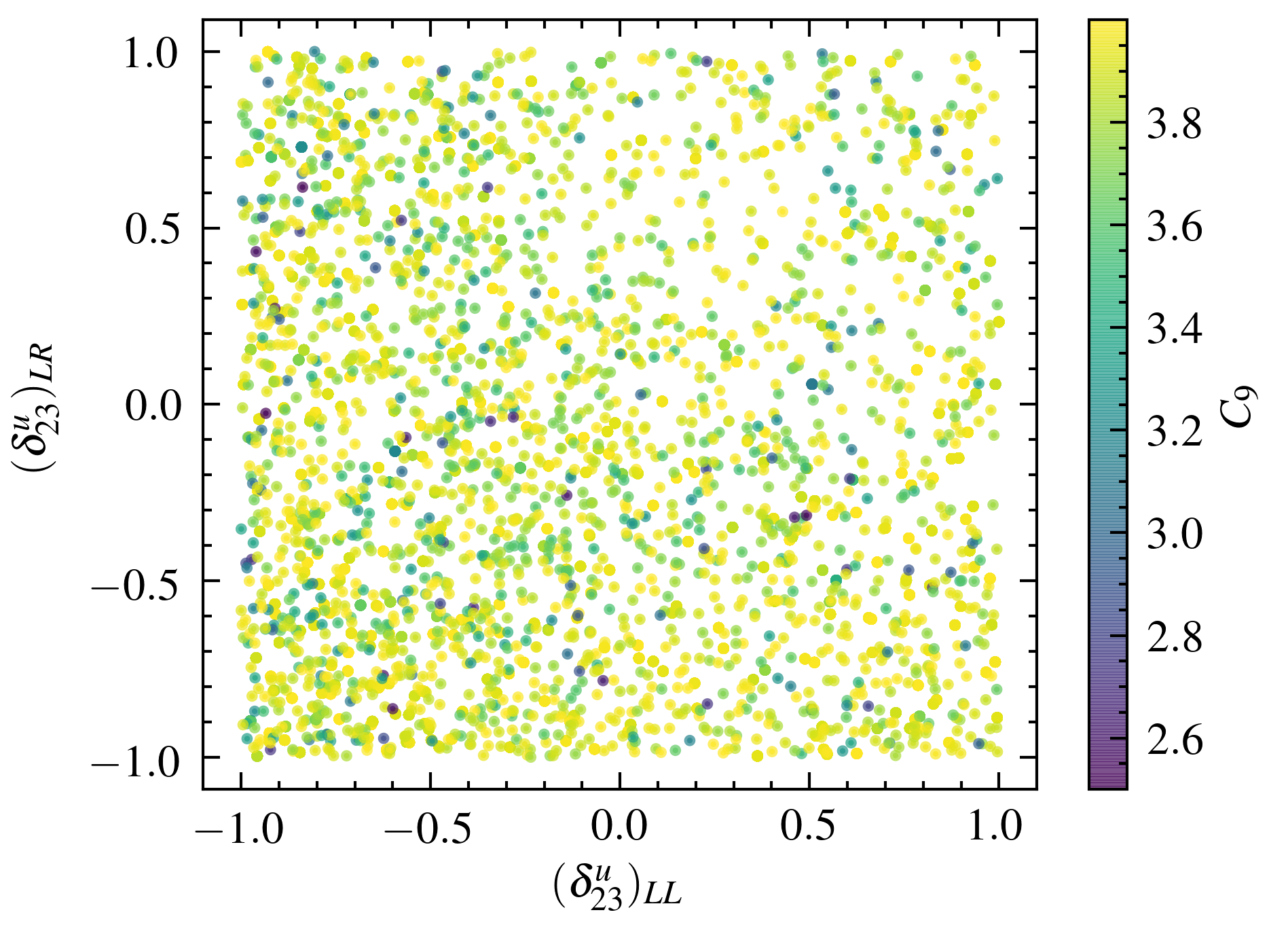}
        \caption{}
        \label{sfig:deltat_uncut}
    \end{subfigure}
    \begin{subfigure}{0.5\textwidth}
        \includegraphics[width=\linewidth]{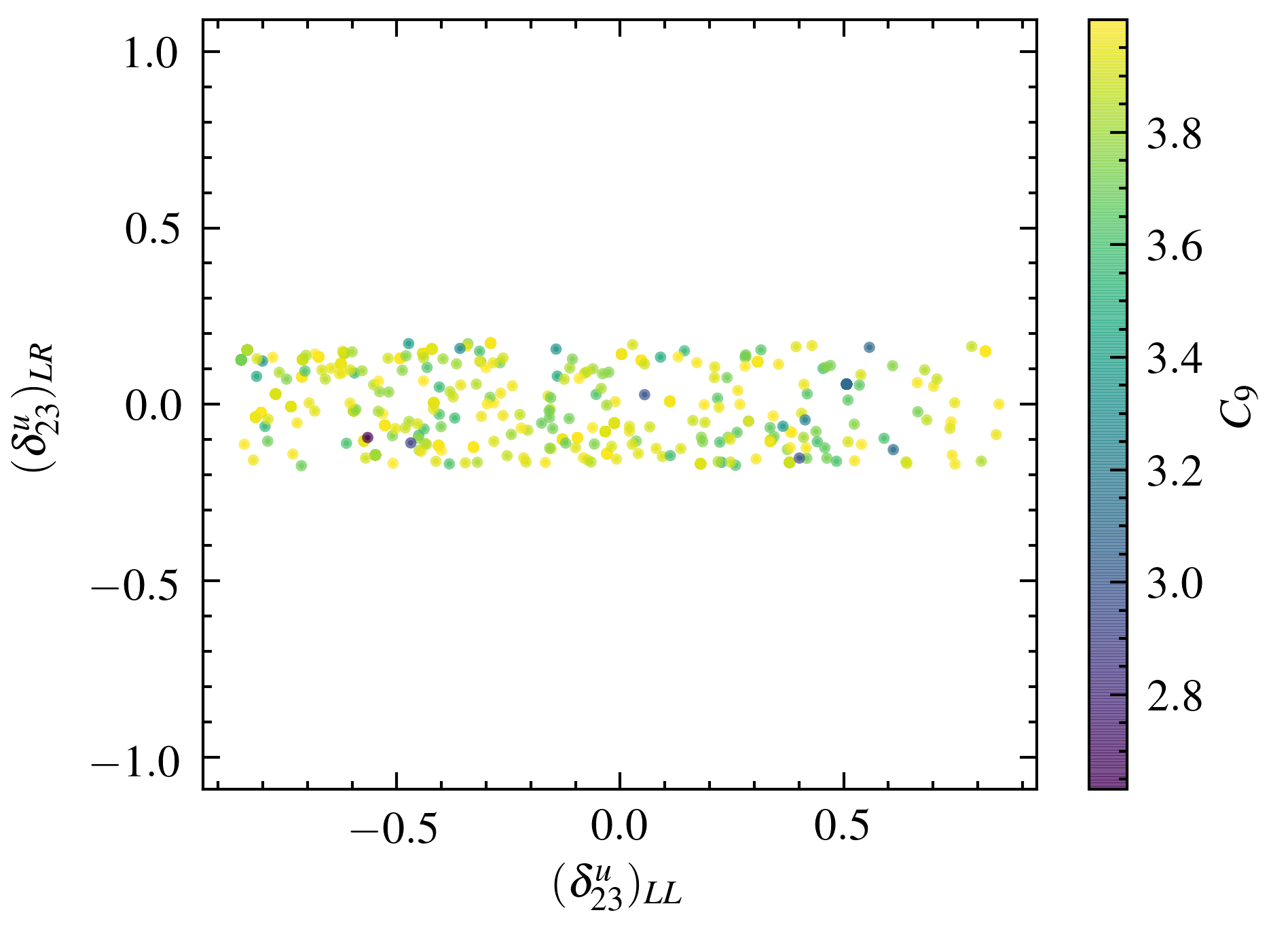}
        \caption{}
        \label{sfig:delta_cut}
    \end{subfigure}
    \caption{Top: distribution of the sampled points in the $(m_{\chi^\pm_1},m_{\chi^0_1})$ plane. First (top-left) in the case of unconstrained model points that significantly shift the value of $C_9$. Applying vacuum stability and tachyon constraints on the $\delta$ parameters yields the top-right plot. The same goes for the bottom plots, in the $\{(\delta^{u}_{23})_{LL},(\delta^{u}_{23})_{LR}\}$ plane.\vspace*{0.5cm}}
    \label{fig:charg_neut_c9c7delta}
\end{figure}

To examine these best-fit points, one can look at the associated mass spectra for some well-studied collider SUSY signals like electroweakinos and coloured sparticle states. \\
In particular, in Fig.~\ref{fig:charg_neut_c9c7delta}, we show both MI parameters and the LSP's mass distribution for our candidate models, without imposing constraints (left) and after imposing constraints (right) for some of the best points with respect to the expected $C_9$ shift. We see that a 10\% fraction survives the tachyonic and vacuum stability constraints while still offering valid candidates for the flavour anomalies.

Similarly to what was shown in Refs.~\cite{Lunghi:1999uk,Behring:2012mv}, it is mostly the top row diagrams in Fig.~\ref{fig:sfigs} corresponding to chargino interactions that contribute the most to $C_9$, which corresponds to $(\delta^u_{23})_{LR}, (\delta^u_{23})_{LL}$ terms in the MIA. For $C_7$, the major contributions come from $(\delta^d_{23})_{LR}$ in the gluino diagrams. However, the  effect on the Wilson coefficients shown here is the result of a global effect coming from all the contributions. The correlation between the free parameters (pMSSM+MI) and the best $(C_9,C_7)$ values was not found to point towards a specific direction. 

The effect of the constraints on the most important MI parameters is also shown in Figs.~\ref{sfig:delta_cut} and~\ref{sfig:deltat_uncut}. From left to right, the available parameter space in the $(\delta_{LL},\delta^u_{LR})$ plane is reduced from $[-1,1]\times[-1,1] $ to $ \simeq [-0.85,0.85]\times[-0.2,0.2]$. This shows that vacuum stability constraints on $\tilde{c}_L-\tilde{t}_R$ mixing are the most stringent ones, as expected from large average squark masses, i.e. $ M_{sq} \gg 2 m_t$. The other MI parameters' contribute very little to $C_{7,9}$ and can be neglected and/or kept close to zero.

The results clearly show the interest of NMFV scenarios, and the need of their further exploration. 
Indeed, the main advantage of the MIA in our case was to easily explore the pMSSM extended with flavour violation, with direct access to the flavour-violating parameters instead of the final mass eigenstates. Also, it has the advantage of reducing the model's free parameters, if their contribution is not significant in the subject at hand, which we did by keeping fewer than 30 parameters, instead of $\mathcal{O}(50)$ or $\mathcal{O}(100)$. However, due to the obviously expected effect on the sparticle spectrum, a more general and complete approach without approximation is necessary to completely confirm the model's shown interesting features. Moreover, a complete approach should also evaluate the contribution of such models to the muon $(g-2)_\mu$. 
This is precisely what is addressed in the next section.

\section{Analytical calculations in NMFV-MSSM scenarios without approximation}\label{sec:NMFV_MARTY}
In the pMSSM, analytical calculations have been performed for several one-loop quantities such as $C_7$, $C_9$~\cite{Bobeth:2004jz} and $(g-2)_\mu$~\cite{Martin:2001st}.
	
In NMFV scenarios, some calculations have been performed at the one-loop level (see e.g. Refs.~\cite{Dedes:2008iw,crivellin2011complete,crivellin2010chirally}), but the general contributions to $C_7$, $C_9$ and $(g-2)_\mu$ are not known. In the following sections, we present the methods that we used to derive analytically these quantities in the general MSSM with $105$ parameters for the first time, together with their evaluation in a particular subset of NMFV scenarios with $42$ parameters.
	
\subsection{Methods}
\subsubsection{Theoretical calculations}
	
\begin{figure}
\centering
 \begin{subfigure}{0.32\textwidth}
    \includegraphics[width=.9\linewidth]{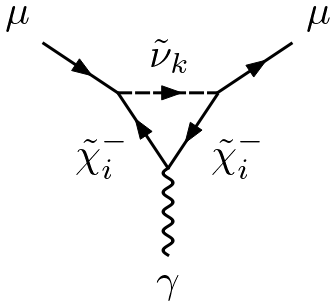}
     \caption{$\tilde{\chi}^+$ penguins in $(g-2)_\mu$}
 \end{subfigure}
  \begin{subfigure}{0.32\textwidth}
    \includegraphics[width=.9\linewidth]{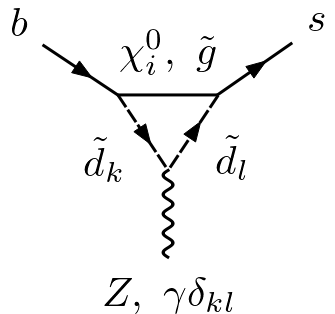} 
    \caption{$\tilde{\chi}^0/\tilde{g}$ penguins in $C_7$ and $C_9$}
 \end{subfigure}
  \begin{subfigure}{0.32\textwidth}
    \includegraphics[width=.9\linewidth]{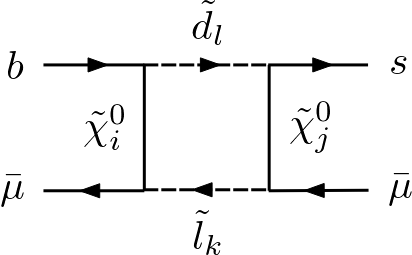}
    \caption{$\tilde{\chi}^0$ boxes in $C_9^\mu$}
    \label{neutralino}
 \end{subfigure}
 	\caption[Feynman diagrams in the NFMV]{Examples of contributions in NMFV-MSSM scenarios. Other chargino, neutralino and Higgs diagrams also contribute to $C_7$, $C_9$ and $(g-2)_\mu$.}
	\label{diagrams}
\end{figure}

In order to derive the full one-loop NMFV contributions to the Wilson coefficients and $(g-2)_\mu$, a large number of Feynman diagrams must be calculated. We performed the analytical calculation in the unconstrained MSSM with general mixings. This means that diagrams must be summed over all particle families: two charginos $\tilde{\chi}^+_{1,2}$, four neutralinos $\tilde{\chi}^0_{1,2,3,4}$, six sleptons $\tilde{l}_{1,2,3,4,5,6}$, six up squarks $\tilde{u}_{1,2,3,4,5,6}$, six down squarks $\tilde{d}_{1,2,3,4,5,6}$ and three sneutrinos $\tilde{\nu}_{1,2,3}$. For the diagram shown in Fig.~\ref{neutralino} for example, there are $4\times 4\times 6\times 6\times 2=1152$ independent diagrams, where the factor of $2$ comes from the two possible contractions for any given ordered pair of neutralinos (counting the crossed diagrams).
	
We used \marty\ \cite{Uhlrich:2020ltd,Uhlrich:2021ded} to calculate automatically all the involved Feynman diagrams and extract the coefficients $(g-2)_\mu$, $C_7$ and $C_9^\mu$. The number of diagrams for each contribution is presented in Table~\ref{contributions}. As \marty\ counts left and right Dirac projectors $P_L$ and $P_R$ as independent vertices, the number of diagrams is larger than what a standard counting method would imply.
\begin{table}[h!]
	\centering
	$\begin{array}{|c|c|c|c|c|c|}
	\hline
	& \tilde{\chi}^+_i & \tilde{\chi}^0_i & \tilde{g} & H^+ & H^0,A^0 \\\hline
	
	(g-2)_\mu    & \pmssm{96} & \pmssm{96} & 0         & \pmssm{1}  & \pmssm{2} \\\hline
	C_7         & \pmssm{240} & \nmfv{96} & \nmfv{24} & \pmssm{24} & 0         \\\hline
	C_9 / \gamma\mathrm{-penguins}   & \pmssm{240} & \nmfv{96}    & \nmfv{24}      & \pmssm{24} & 0 \\\hline
	C_9 / Z\mathrm{-penguins}        & \pmssm{624} & \nmfv{1344}  & \nmfv{240}     & \pmssm{78} & 0 \\\hline
	C_9 / \mathrm{boxes}             & \pmssm{864} & \nmfv{13824} & 0              & \pmssm{12} & 0 \\\hline
	\end{array}$
	\caption[List of contributions in the NMFV]{Number of diagrams for each contribution calculated by \marty. NMFV-specific contributions are the starred numbers. By definition, $C_7$ and $(g-2)_\mu$ only receive contributions from $\gamma$-penguin diagrams. There are in total $17949$ Feynman diagrams.}
	\label{contributions}
\end{table}
	
\subsubsection{Numerical evaluation}
	
The mathematical expressions resulting from the sum of thousands of one-loop diagrams are too large for any analytical purpose. In order to obtain predictions, \marty\ generates a numerical C++ library containing functions evaluating the results given a general MSSM scenario. From a set of values for the SUSY-breaking parameters presented in Eqs.~(\ref{eq:Lsoft_gaugino}) to (\ref{eq:LSOFt:tril}), we are therefore able to evaluate the exact values of $C_7$, $C_9^\mu$ and $(g-2)_\mu$ at the one-loop level in the library generated by \marty.
	
While \marty\ also generates a tree-level spectrum generator to calculate masses and mixings from the initial model parameters, loop corrections are known to be large and we therefore use \Verb+SPheno+~\cite{Porod:2003um, Porod:2011nf} to produce a more precise spectrum including loop-level corrections and phenomenological constraints. Finally, the values of the Wilson coefficients are given to \Verb+SuperIso+ to apply renormalization group equations and evolve the coefficients down to the $b$ mass scale, and calculate flavour observables.
	
\subsubsection{Random scan}
To sample the MSSM parameter space, we used a uniform random scan in $42$ dimensions with NMFV only in the squark sector to reduce the number of free parameters. Input parameter ranges are presented in table~\ref{tab:inputparam}. 
     
\begin{table}[h!]
	\centering
	$\begin{array}{@{}ll@{}}
	\toprule
	\text{Parameter} & \text{Scanned range}\\ \midrule
	\tan\beta     & [2, 60] \\
	\mu     & [-100, 1000]\GeV \\
	M_1,M_2 & [100, 3000]\GeV\\
	M_3 & [100, 7000]\GeV\\
	M_A & [100, 5000]\GeV\\
	(M_{ \tilde{Q}}^2)_{ii} & [10^2,10^7]\GeV^2\\
	(M_{ \tilde{U}}^2)_{ii} & [10^2,10^7]\GeV^2\\
	(M_{ \tilde{D}}^2)_{ii} & [10^2,10^7]\GeV^2\\
	(M_{\tilde{L}}^2)_{ii} & [10^2,10^6]\GeV^2\\
	(M_{ \tilde{E}}^2)_{ii} & [10^2,10^5]\GeV^2\\
		(A_e)_{33}     & [-100, 100]\GeV \\
		(A_{u/d})_{11} & [-0.1, 0.1]\GeV \\
		(A_{u/d})_{22} & [-100, 100]\GeV \\
		(A_{u/d})_{33} & [-10^{4}, 10^4]\GeV \\
		(M_{\tilde{Q}}^2)_{23} & [0,10^3]\GeV^2\\
		(M_{ \tilde{D}}^2)_{23} & [0,10^3]\GeV^2\\
		(A_u)_{ij},i\neq j & [-100, 100]\GeV\\
		(A_d)_{ij},i\neq j & [-100, 100]\GeV \\ \bottomrule
	\end{array}$ 

	\caption{Input parameters for the scan. Specific ranges have been chosen empirically to improve the scan efficiency. There are in total $42$ free parameters, which include the $19$ pMSSM parameters, $14$ flavour violating parameters  $(M_{\tilde{Q}}^2)_{23}$, $(M_{ \tilde{D}}^2)_{23}$, $(A_u)_{ij}$ and $(A_d)_{ij}$ for $i\neq j$.}
	\label{tab:inputparam}
\end{table}
	
\begin{figure}
	\centering
	\includegraphics[width=0.32\linewidth]{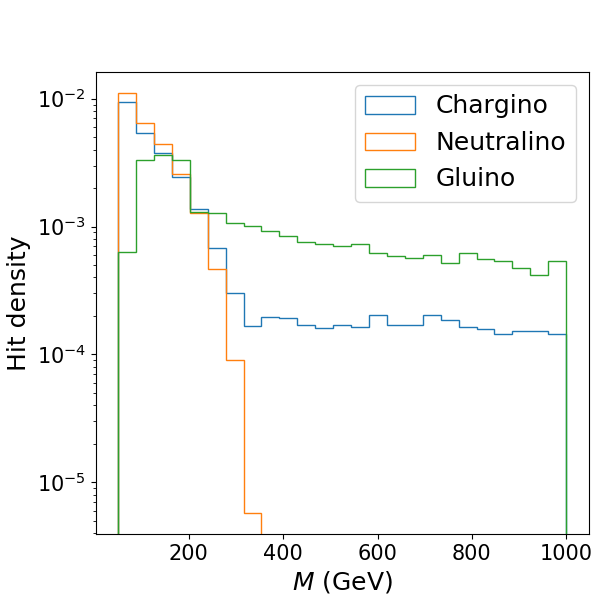}~~
	\includegraphics[width=0.32\linewidth]{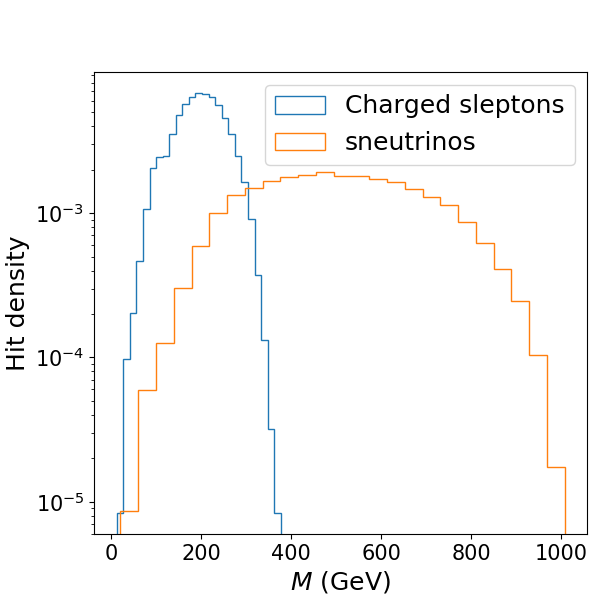}~~
	\includegraphics[width=0.32\linewidth]{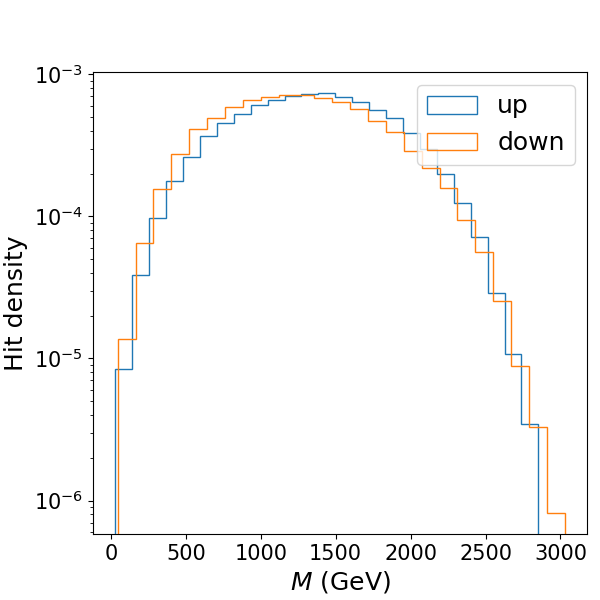}
	\caption{Posterior distributions for gaugino (left), slepton (middle) and squark (right) masses. For particle families, the distribution corresponds to the lightest particle of the family. Chargino and gluino mass distributions extend up to $3\TeV$ and $7\TeV$ respectively.}
	\label{posteriors}
\end{figure}

The scan efficiency is of about $0.05\%$, corresponding to physical scenarios for which \Verb+SPheno+ can calculate a spectrum. For such a low efficiency there is a large bias in the selected scenarios. Consequently, we also present some posterior distributions of the spectrum in Fig.~\ref{posteriors}. The scan could be refined with better constraints on the input parameters to improve the efficiency. The following analysis is therefore more a proof of principle rather than a complete phenomenological study of the MSSM parameter space. 
There are two visible biases in the posterior distributions of spectrum parameters:
\begin{itemize}
	\item Charged sleptons are lighter than sneutrinos because the range for $M_{ \tilde{E}}^2$ is smaller than that of $M_{\tilde{L}}^2$. 
	\item The lightest neutralino is always lighter than $400\GeV$ contrary to the lightest chargino. This is because we impose the condition of having a neutral LSP in order to be a dark matter candidate. 
\end{itemize}
	
To improve the scan efficiency, we considered machine learning techniques to sample the parameter space. The purpose of these techniques is to create a sampling bias toward scenarios that generate valid model points, that therefore improves the scan efficiency. However, while these techniques can be implemented without much difficulty for the pMSSM with $19$ parameters, the $43$-dimensional space of the NMFV scenarios we present in this paper is too large for the machine-learning-based sampling to be established. Indeed, in the absence of prior knowledge on the distribution of valid parameters, and because of the high number of dimensions, no efficient sampler could  be constructed with the considered techniques such as Normalizing Flows or Hamiltonian Monte Carlo samplers. Further work is required to build efficient samplers in highly dimensional unknown and little-constrained parameter spaces with very few acceptable points, which is beyond the scope of this work. 
	
Finally, let us stress that the current LHC limits on SUSY particle masses are not directly applicable to our study. The NMFV-MSSM being a more general model than the so-called simplified or constrained MSSM scenarios, the recasting of collider constraints on the sparticle spectrum is a non-trivial task (see e.g. Refs.~\cite{Bernigaud:2018vmh,Alguero:2021dig} and yields weaker bounds. We nevertheless checked for points leading to significant negative contributions to $C_9$ that they escape the direct limits, in particular due to the degeneracy between the lightest neutralino and charigno $\Delta m(\chi^0_1,\chi^\pm_1) \leq 1~{\rm GeV}$, which makes them extremely complicated to probe experimentally. 
	
\subsection{Results}
	
Using as input the NMFV-MSSM spectra obtained with \Verb+SPheno+, the numerical functions generated by \marty\ evaluate the full one-loop contributions to the Wilson coefficients and $(g-2)_\mu$. As the scan is random, we show distributions for the different quantities that we calculated for the $70282$ valid model points. In the following, we study the impact on the Wilson coefficients and $(g-2)_\mu$ separately. Then, the relation between the two will be discussed.
	
\subsubsection{Wilson coefficients}
	
\begin{figure}[h!]
	\centering
	\includegraphics[width=0.45\linewidth]{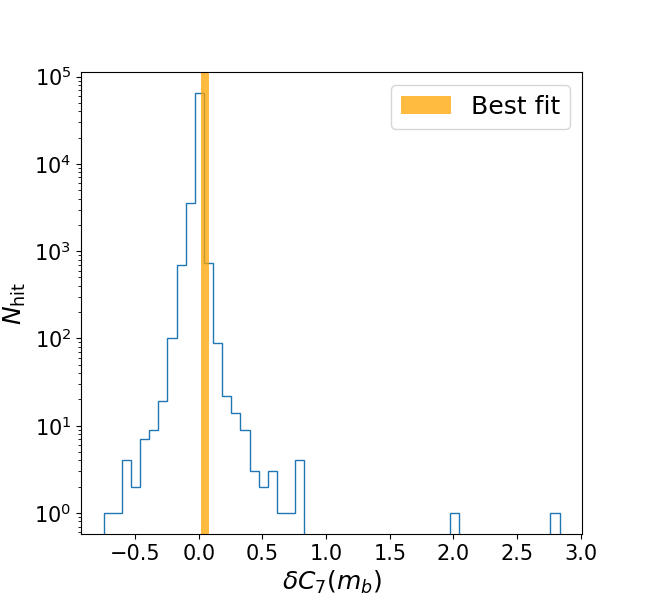}
	\hspace{1cm}
	\includegraphics[width=0.445\linewidth]{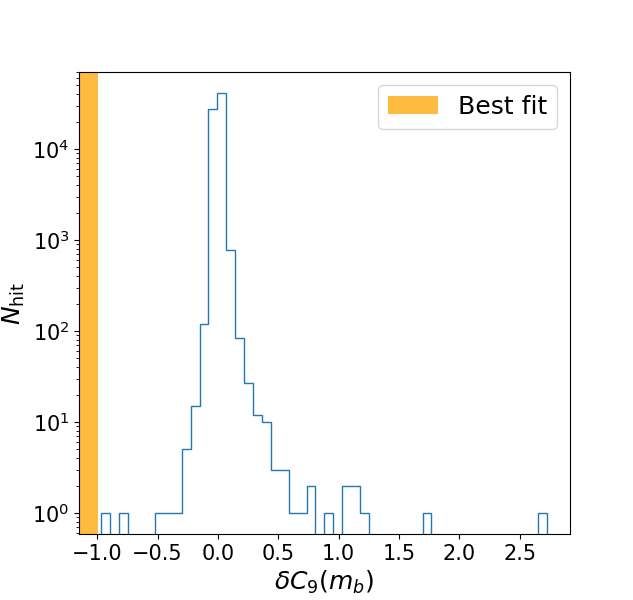}
	\caption[Distribution of $C_7$ and $C_9$]{\label{distribwilson}Distribution of the Wilson coefficients $\delta C_7$ and $\delta C_9^\mu$. The $1\sigma$ best-fit regions from Ref.~\cite{Hurth:2021nsi} are shown in orange.}
\end{figure}

The distributions for the NMFV-MSSM contributions to the Wilson coefficients $C_7$ and $C_9^\mu$ are presented in Fig.~\ref{distribwilson}.

Both distributions are centered around zero, as expected. While the majority of $\delta C_7$ points are close to zero and the best-fit region, many scenarios are already excluded because of a large shift to this coefficient. For $\delta C_9^\mu$, the best fit-region is shifted by $-1$ from the SM value. While it is possible to obtain substantial $C_9$ shifts in our scenarios, only a handful of them predict $\delta C_9^\mu< 0.2$.
	
It is important to note that the best-fit region for $C_9^\mu$ should not be considered as a discriminant criterion, any scenario between the SM and the best fit can still fit better flavour observables and should be carefully considered. 
	
\begin{figure}[h!]
	\centering
	\includegraphics[width=0.45\linewidth]{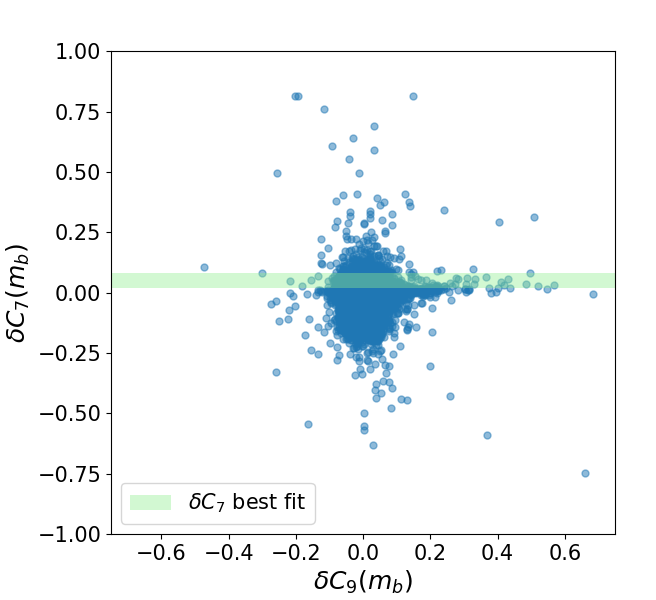}
	\caption[Combined distribution of $C_7$ and $C_9$]{\label{distrib79}Combined distribution of the Wilson coefficients $\delta C_7$ and $\delta C_9^\mu$. The best-fit region for $\delta C_7$ is shown in green.}
\end{figure}

A $2D$ distribution of $(\delta C_7,\delta C_9^\mu)$ is presented in Fig.~\ref{distrib79}. It is clear that the constraint on $\delta C_7$ excludes several scenarios with $\delta C_9^\mu < -0.15$. It seems nevertheless possible to address both coefficients, but a larger dataset is required to explore the region with large negative $\delta C_9$. 

\subsubsection{$\mathbf{(g-2)_\mu}$ and combined analysis}\label{sec:SLHA2_combined_analysis}	

For just over fifteen years, the anomalous magnetic moment of the muon has proven to be a persistent tension between the SM~\cite{Aoyama:2020ynm,Aoyama:2012wk,Aoyama:2019ryr,Czarnecki:2002nt,Gnendiger:2013pva,Davier:2017zfy,Keshavarzi:2018mgv,Colangelo:2018mtw,Hoferichter:2019mqg,Davier:2019can,Keshavarzi:2019abf,Kurz:2014wya,Melnikov:2003xd,Masjuan:2017tvw,Colangelo:2017fiz,Hoferichter:2018kwz,Gerardin:2019vio,Bijnens:2019ghy,Colangelo:2019uex,Blum:2019ugy,Colangelo:2014qya} and experimental measurements. The most recent results obtained at Fermilab~\cite{Muong-2:2021ojo} have not only confirmed the Brookhaven $3\sigma-4\sigma$~\cite{Muong-2:2006rrc} discrepancy, but raised it to the $4.2\sigma$ level with a combined experimental average of $a_\mu^{\rm EXP}=116592061(41) \times 10^{-11}$. However, multiple questions remain as lattice QCD calculations may reduce the discrepancy to only $1.6\sigma$~\cite{Borsanyi:2020mff}. In the following, we investigate whether NMFV-MSSM models can account for the observed tensions in both $(g-2)_\mu$ and the $b$ quark flavour sector.

\begin{figure}[h!]
	\centering
	\includegraphics[width=0.5\linewidth]{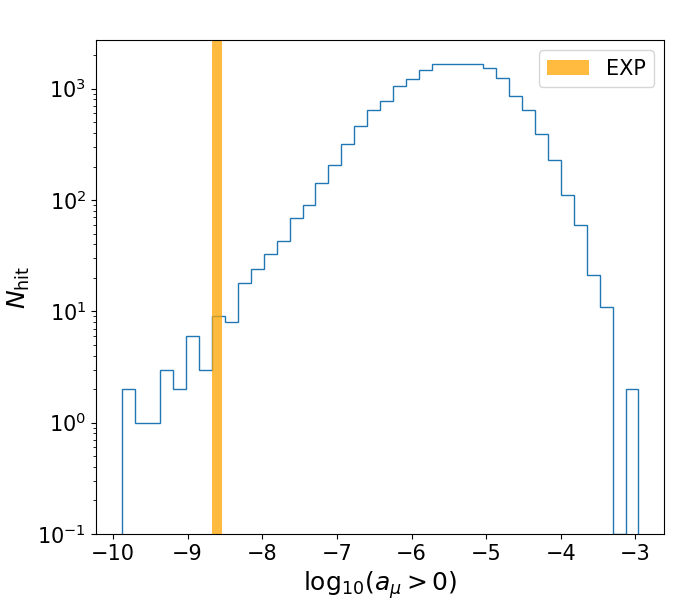}
	\caption[Distribution of $(g-2)_\mu$]{\label{distribgm2}Distribution of $\delta(g-2)_\mu$. Only scenarios with a positive shift are considered and the experimental measurement with its $1\sigma$ uncertainty~\cite{Muong-2:2021ojo} is shown in orange.}
\end{figure}

Our present analysis does not strictly consider NMFV parameters in the lepton sector\footnote{There is no limitation for NMFV in the lepton sector, this choice has been made to reduce the number of free parameters and concentrate on flavour observables that are more difficult to address because of the $C_9^\mu$ shift.} as shown in Table~\ref{tab:inputparam}. We present the numerical results for $(g-2)_\mu$ in the following. The mass distribution for charged sleptons is around the electroweak scale, i.e. a few hundred $\GeV$ (see Fig.~\ref{posteriors}). This implies significant contributions to $(g-2)_\mu$ that are shown in Fig.~\ref{distribgm2}. As the experimental deviation is very small~\cite{Muong-2:2021ojo}, it is not hard to address $(g-2)_\mu$ alone. 
	
\begin{figure}[h!]
	\centering
	\includegraphics[width=0.5\linewidth]{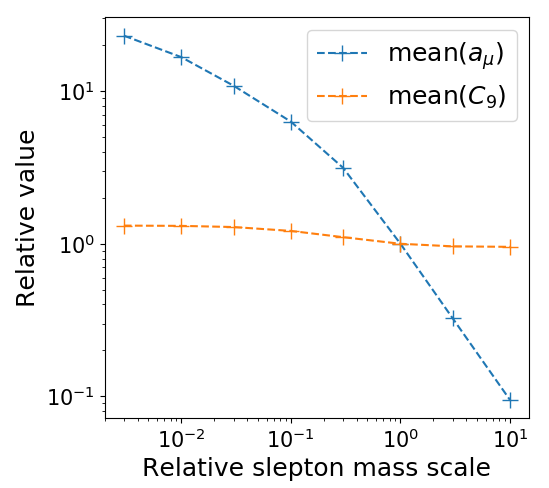}
	\caption[Coupling to the slepton mass scale]{The variation of the relative mean absolute value of $C_9$ and $(g-2)_\mu$ for the entire dataset is plotted as a function of the relative slepton mass scale. The initial, non-modified dataset corresponds to the point at $(1,1)$.}
	\label{decoupling}
\end{figure}

As shown in table~\ref{contributions}, the lepton and quark sectors are sensitive to the neutralino and chargino mass scales. However, while there are slepton contributions in box diagrams for $C_9^\mu$, these contributions are small and the latter coefficient is almost independent of the slepton masses. Figure~\ref{decoupling} shows the dependence of $(g-2)_\mu$ and $C_9^\mu$ with respect to the relative slepton mass scale.\footnote{$C_7$ is completely independent of the slepton sector.} 
	
This analysis shows that by rescaling the slepton masses (charged sleptons and sneutrinos), one can shift the value of $(g-2)_\mu$ and let Wilson coefficients $C_7$ and $C_9^\mu$ remain stable. It is therefore possible to search for a scenario that fits the flavour observables well and adjust the slepton mass scale to address $(g-2)_\mu$.

\section{Conclusion}\label{sec:conclusion}
We presented a first study of the pMSSM extended with non-minimal flavour-violating couplings in the context of the tensions observed in $b\to s \ell^+\ell^-$ transitions with the SM predictions, and considered the SUSY contributions to the relevant Wilson coefficients. 
We carry on our study first by assuming the mass insertion approximation and show that the NMFV contributions allow us to shift the Wilson coefficients sufficiently enough to fully address the anomalies. After imposing theoretical constraints on the flavour-violating parameters we still find scenarios in agreement with the experimental measurements.
     
While the MIA provides a direct access to the flavour-violating parameters and eases the phenomenological studies by reducing the number of free parameters, a more general approach is necessary to fully assess the impact of NMFV contributions. Hence, in a second part, we calculated for the first time the full one-loop analytical contributions in the general MSSM to the relevant Wilson coefficients, as well as to $(g-2)_\mu$ using \marty. By scanning the MSSM parameter space randomly and setting non-zero values for some of the flavour-violating parameters, we obtained 70282 valid scenarios with their individual spectra. In these scenarios, we showed that $C_9^\mu$ can be shifted towards the best-fit region given in Ref.~\cite{Hurth:2021nsi} but that we have only a few points that shift $C_9^\mu$ in the favoured direction and let $C_7$ come close to the SM prediction. We then discussed the scaling of $(g-2)_\mu$ with the slepton mass scale that allows us to address $(g-2)_\mu$ without modifying the predictions for flavour observables.
    	
The present analysis is limited by the small sample of scenarios. As a perspective, the scan should be optimized by searching a parameter set that is more likely to produce physical scenarios. In particular, by looking at the posterior distributions of the input parameters it is possible to refine the scan, improve the efficiency and generate more scenarios to analyse. Finally, experimental constraints could be studied more in depth to compare the obtained spectra with direct searches of SUSY particles, in particular from LHC measurements. The obtained results are nevertheless very promising and show for the first time the impact of NMFV parameters in addressing the recent anomalies. 

\section*{Acknowledgements}
   We would like to thank Gérald Grenier, Sabine Kraml, Gaël Alguero, Björn Herrmann and Abdelhamid Boussejra for helpful discussions and insights. 
 \newpage
\appendix 
\section{Wilson coefficients}\label{sec:A_wilson_coeff}
    
\subsection{Effective Hamiltonian at the electroweak scale}

The effective Hamiltonian for the decay $ B \rightarrow X_s \ell^+\ell^-$
in the SM and in the MSSM is given by (neglecting the
small contribution proportional to $V^*_{us} V_{ub}$), in the basis of Ref. \cite{Lunghi:1999uk}

\begin{equation}
{\cal H}_{\rm eff} = - \frac{4 G_F}{\sqrt{2}} V^*_{ts} V_{tb} \left[
                     \sum_{i=1}^8 C_i (\mu) O_i + 
                     \frac{\alpha}{4 \pi} \sum_{i=9}^{10} \tilde{C}_i (\mu) O_i
                     \right]\ ,
\end{equation}
where the $O_i$ operators read
\begin{align*}
    O_1 &= \overline{s}_{L\alpha} \g_\mu b_{L\alpha } \overline{c}_{L\beta} \g^\mu c_{L\beta}\ , \\ 
    O_2 &= \overline{s}_{L\alpha} \g_\mu b_{L\beta } \overline{c}_{L\beta} \g^\mu c_{L\alpha}\ , \\ 
    O_3 &= \overline{s}_{L\alpha} \g_\mu b_{L\alpha } \sum_{q=u,..,b} \overline{q}_{L\beta} \g^\mu q_{L\beta}\ , \\ 
    O_4 &= \overline{s}_{L\alpha} \g_\mu b_{L\beta } \sum_{q=u,..,b} \overline{q}_{L\beta} \g^\mu q_{L\alpha}\ , \\ 
    O_5 &= \overline{s}_{L\alpha} \g_\mu b_{L\alpha } \sum_{q=u,..,b} \overline{q}_{R\beta} \g^\mu q_{R\beta}\ , \\ 
    O_6 &= \overline{s}_{L\alpha} \g_\mu b_{L\beta } \sum_{q=u,..,b} \overline{q}_{R\beta} \g^\mu q_{R\alpha}\ , \\ 
    O_7 &=\frac{e}{16 \pi^2}m_b\overline{s}_L\sigma^{\mu\nu}b_R F_{\mu\nu}\ , \\
    O_8 &=\frac{g_s}{16 \pi^2}m_b \overline{s}_L T^a\sigma^{\mu\nu}b_R G_{\mu\nu}^a\ , \\
    O_9 &=(\overline{s}_L\g_\mu b_L)\overline{l}\g^\mu l\ , \\
    O_{10} &=(\overline{s}_L\g_\mu b_L)\overline{l}\g^\mu \g_5 l\ , 
\end{align*}
with $V$ being the CKM matrix and $\ds q_{L(R)}= \frac{(1\mp \Gc)}{2} \;q$. \\
This Hamiltonian is known to next-to-leading order both in the SM~\cite{Buras:1994dj,Buchalla:1995vs} and
in the MSSM~\cite{Cho:1996we,Goto:1996dh,Goto:1998qv}. \\
For the $B$ system, the operators and coefficients of interest for the anomalies are $C_7, C_9,$ and $C_{10}$.

\subsection{Wilson coefficients}\label{sec:wilson_coeff}
We consider the contributions to the Wilson coefficients as given in Ref.~\cite{Lunghi:1999uk}, including the correction as suggested in Ref.~\cite{Behring:2012mv}, which we reproduce here for completeness. 
The loop functions $P_{ijk}(a,b)$ are defined in Appendix~\ref{sec:hypgeo}. In what follows, the constants are
\begin{itemize}
\item[--] $ \theta_W$ is the weak mixing angle of the SM for the electroweak bosons.\\[-0.8cm] 
\item[--] $M_W^2$ is the squared mass of the $W^\pm$ boson.\\[-0.8cm] 
\item[--] $ \lambda_{t,b}$ are the Yukawa couplings for the top and bottom quarks.\\[-0.8cm]  
\item[--] $ g_2$ is the weak isospin coupling constant.\\[-0.8cm]  
\item[--] $M_{sq}$ is the average squark mass. \\[-0.8cm] 
\item[--] $M_{\tilde{g}}$ is the gluino mass. \\[-0.8cm] 
\item[--] In $P_{ijk}(x_i,x_j)$, $x_i, x_j$ are defined as $M_{\tilde{\chi}_i}^2 / M_{sq}^2$. for chargino loops.\\[-0.8cm] 
\item[--] In $P_{ijk}(x,x)$, $x$ is defined as $M_{\tilde{g}}^2 / M_{sq}^2$, for gluino graphs. \\[-0.8cm] 
\item[--] The $V_{ab}$ are the elements of the CKM matrix.\\[-0.8cm] 
\item[--] $m_{s,b}$ are the pole masses of the $s,b$ quarks.\\[-0.8cm] 
\item[--] $N_c = 3$ is the number of color charges in $SU(3)_c$.\\[-0.8cm] 
\item[--] $\alpha_s$ is the QCD coupling constant, evaluated at $M_Z$.\\[-0.8cm] 
\item[--] $G_F$ is the Fermi constant. \\[-0.8cm] 
\item[--] $U,V$ are the charginos mixing matrices defined in eq.\eqref{eq:char2}.
\end{itemize}

In the weak eigenstates basis the chargino mass matrix is given by \cite{drees2005theory}:
\begin{equation}
M_\chi =\left(
\begin{array}{cc}
M_2 & \sqrt{2}M_W \sin \beta \\
 \sqrt{2}M_W \cos \beta &\mu
\end{array}
\right) \ ,
\label{eq:char2}
\end{equation}
where the index 1 of rows and columns refers to the wino, and the index 2 to the higgsino. $\mu$ is the Higgs quadratic coupling and $M_2$ the soft SUSY breaking wino mass. The $3\times 3$ complex matrices $U$ and $V$ which diagonalize $M_{\chi}$ are introduced:
\begin{equation}
{\rm diag}(M_{\chi_1},M_{\chi_2})= U^* M_{\chi} V^+  \ .  
\end{equation}
Their explicit expressions can be found in e.g. Ref.~\cite{kruger2000flavor}. 
All the contributions to the Wilson coefficients are evaluated at the renormalization scale $\mu_0 = m_W$.

\subsubsection{Chargino contributions}
In the following, we give the contributions from chargino loops, such as the two top diagrams in Fig.~\ref{fig:sfigs}. \\ 

\textbf{Z-penguin with higgsino/wino loops:}
\begin{equation}
    \begin{aligned}
        &- \frac{C_9}{1-4\sw}=C_{10}=  
        (\delta_{23}^u)_{LR}  \frac{\lambda_t}{g_2}         \frac{V_{cs}^*}{V_{ts}^*}
        \frac{1}{4 \sw}
        \sum_{i,j=1,2} V_{i1} V_{j2}^* \times   \\ 
        & \left\{U^*_{i1} U_{j1} \sqrt{x_{i} x_{j}}     P_{112}(x_i,x_j) 
        + V^*_{i1} V_{j1} P_{111}(x_i,x_j)-
        \frac{1}{2} \delta_{ij} P_{021}(x_i,x_j)
        \right\} \ .
    \end{aligned}
\end{equation}
This diagram is proportional to $(\delta^u_{23})_{LR}$, which is one of the most interesting mass insertions, and is yet to be more constrained. \\

\textbf{Z-penguin with two wino vertices:} 
\begin{equation}
        \begin{aligned}
            &-\frac{C_9}{1-4\sw}=C_{10}= 
            -(\delta_{23}^u)_{LL} {V_{cs}^*\over V_{ts}^*}{1\over 4 \sin^2{\theta_W}}
             \sum_{i,j=1,2} V_{i1} V_{j1}^* \times   \\
             &
            \left\{
            U^*_{i1} U_{j1}
            \sqrt{x_{i} x_{j}} P_{112}(x_i,x_j) 
             + V^*_{i1} V_{j1} P_{111}(x_i,x_j)-
             \delta_{ij} P_{021}(x_i,x_j)
            \right\} \ .
            \label{eq:winw}
        \end{aligned}
\end{equation}
This is the same diagram as above, but with the exchange of two winos. They differ only by the specific mass insertion and the  factor $\lambda_t/g_2$. Both diagrams are null in the limit of a diagonal chargino mass matrix, and so they are negligible for large $M_2$.\\

\textbf{Gamma penguin with two wino vertices}: 

    \begin{align}
        C_7=& -(\delta^u_{23})_{LL} \frac{M^2_W}{M_{sq}^2 }\frac{1}{3} {V_{cs}^* \over V_{ts}^*} 
        \sum_{i=1,2} V_{i1} V^*_{i1} \left\{\frac{3}{2} P_{222}(x_i,x_i)+
        P_{132}(x_i,x_i)\right\} \ ,         \\
        C_9=& -(\delta^u_{23})_{LL}  \frac{M^2_W}{M_{sq}^2 }\frac{2}{3} {V_{cs}^*\over V_{ts}^* }\times   \sum_{i=1,2} V_{i1} V^*_{i1} \left\{ P_{312}(x_i,x_i)- \frac{1}{3} P_{042}(x_i,x_i)+ x_i P_{313}(x_i,x_i) \right\}  \ ,  \\
        C'_7=&-(\delta^u_{23})_{LL} \frac{M^2_W}{M_{sq}^2 }\frac{1}{3} {V_{cs}^* \over V_{ts}^*} \frac{m_s}{m_b}
        \sum_{i=1,2} V_{i1} V^*_{i1} \left\{\frac{3}{2} P_{222}(x_i,x_i)+ P_{132}(x_i,x_i)\right\}\ .
    \end{align}
    
\textbf{Gamma penguin with higgsino-wino vertex: }
      \begin{align}
        C_7=& \frac{M^2_W}{M_{sq}^2 }  {V_{cs}^* \over V_{ts}^*} \sum_{i=1,2} 
        \left[
        V_{i2}^* V_{i1}  {\lambda_t \over g_2} 
        \left\{\frac{1}{2} P_{222}(x_i,x_i)+{1\over 3} P_{132}(x_i,x_i)\right\} 
        (\delta^u_{23})_{LR} + \right.   \nonumber \\
        &  \left. U_{i2}  V_{i1} {M_{\chi_i} \over m_b} {\lambda_b \over g_2} 
        \left\{P_{212}(x_i,x_i)+{2\over 3} P_{122}(x_i,x_i)\right\} 
        (\delta^u_{23})_{LL} \right] \ ,    \\
        C_9=&(\delta^u_{23})_{LR}  \frac{M^2_W}{M_{sq}^2 }\frac{2}{3} {\lambda_t \over g_2}{V_{cs}^*\over V_{ts}^* } \sum_{i=1,2} V_{i2}^* V_{i1} 
        \left\{ P_{312}(x_i,x_i)- \frac{1}{3} P_{042}(x_i,x_i)+ x_i P_{313}(x_i,x_i) \right\} \ ,
          \\
        C'_7=&(\delta^u_{23})_{LR} \frac{M^2_W}{M_{sq}^2 }\frac{1}{3} {\lambda_t \over g_2}   {V_{cs}^* \over V_{ts}^*}
        \frac{m_s}{m_b} 
        \sum_{i=1,2} V_{i2}^* V_{i1} \left\{\frac{3}{2} P_{222}(x_i,x_i)+ P_{132}(x_i,x_i)\right\}\ .
        \end{align}
The primed operators are obtained by switching the chirality of external states. All of these contributions are used, but we are most interested in the $C_9$ contribution coming from the $\gamma \tilde{H} \tilde{W}$ vertex. \\

\textbf{Z-penguin with two wino vertices and a double mass insertion:} \\
Even though a double mass insertion corresponds to a higher order in the perturbative expansion,  it has been pointed out \cite{Lunghi:1999uk} that this particular diagram could provide enhancement in the $K$ system. For completeness, in the $B$ decay, the contribution is: 
\begin{equation}
     \begin{aligned} 
            & -\frac{C_9}{1-4\sw}=C_{10}=-{ (\delta_{23}^u)_{LR} (\delta_{33}^u)_{LR} \over 4 \sw} 
            {V_{cs}^* \over V_{ts}^*} \sum_{i,j=1,2} V_{i1} V_{j1}^* \times   \\
            & \left\{
            U^*_{i1} U_{j1} \sqrt{x_{i} x_{j}} P_{123}(x_i,x_j) +\frac{1}{2} V^*_{i1} V_{j1} 
            P_{122}(x_i,x_j)- \frac{\delta_{ij}}{3} P_{032}(x_i,x_j)
            \right\}  \ .
            \label{eq:WWZ-2MI} 
        \end{aligned}
\end{equation}     
\subsubsection{Gluino contribution}
In this part, we collect all the contributions arising from gluino loops, from the bottom diagrams in Fig.~\ref{fig:sfigs}, with $x \equiv M_{\tilde{g}}^2/M^2_{sq}$ . \\

\textbf{$\gamma$-penguin:}
        \begin{align}
            C_7&=\frac{\sqrt{2}}{M_{sq}^2 G_F} \ \frac{1}{3}\ \frac{N_c^2-1}{2N_c}
             {\pi \as \over V_{ts}^* V_{tb}}  
            \left[\left( (\delta^d_{23})_{LL}+(\delta^d_{23})_{RR} \frac{m_s}{m_b}\right) 
                   \frac{1}{4}P_{132}(x,x)+ (\delta^{d}_{23})_{RL} P_{122}(x,x) \frac{M_{\tilde{g}}}{m_b}\right] \ ,
               \\
            C_7^\prime&= \frac{\sqrt{2}}{M_{sq}^2 G_F}\ \frac{1}{3}\ \frac{N_c^2-1}{2N_c}
              {\pi \as \over V_{ts}^* V_{tb}}  
            \left[\left( (\delta^d_{23})_{RR}+(\delta^d_{23})_{LL} \frac{m_s}{m_b}\right) 
                   \frac{1}{4}P_{132}(x,x)+ (\delta^{d}_{23})_{LR} P_{122}(x,x) \frac{M_{\tilde{g}}}{m_b}\right] \ ,
              \\
            C_9 &=
            -\frac{\sqrt{2}}{M_{sq}^2 G_F}\ \frac{1}{3} \  \frac{N_c^2-1}{2N_c}
            {\pi \as \over V_{ts}^* V_{tb}}   \frac{1}{3} 
            P_{042}(x,x) (\delta^d_{23})_{LL} \ ,
               \\
            C_9^\prime &=
            -\frac{\sqrt{2}}{M_{sq}^2 G_F}\ \frac{1}{3}\  \frac{N_c^2-1}{2N_c} 
            {\pi \as \over V_{ts}^* V_{tb}}  \frac{1}{3} P_{042}(x,x) 
            (\delta^d_{23})_{RR} \ .
            \label{eq:glu}
        \end{align}
The terms proportional to $M_{\tilde{g}}$ can be dominant over the others. However, the mass insertion which enters the diagram is strongly constrained from $b \to s \gamma$ \cite{Gabbiani:1988rb}. \\ 

\textbf{Double Mass Insertion  -- $Z-\tilde{g}$}: \\
For completeness, we give the gluino penguin contribution with a double mass insertion:
    \begin{align}
    &&-\frac{C_9}{1-4\sw}=C_{10} = \frac{(\d^d_{33})_{LR}(\d^d_{23})_{RL}}{V_{tb} V_{ts}^*} 
    \frac{N_c^2-1}{2N_c} \frac{\as}{12  \a } \ P_{032}(x,x)\   \ , \\
    &&-\frac{C'_9}{1-4\sw}=C'_{10} = \frac{(\d^d_{33})_{RL}(\d^d_{23})_{LR}}{V_{tb} V_{ts}^*}
    \frac{N_c^2-1}{2N_c} \frac{\as}{12  \a} \ P_{122}(x,x) \ .
    \label{eq:gluz2}
    \end{align}
\subsubsection{Box diagrams}
The following contributions come from the box diagrams in Fig~\ref{fig:box}. \\
\\
\textbf{Box diagram with wino exchange:}
        \begin{equation}
            C_9 = - C_{10} = (\delta^u_{23})_{LL} {V_{cs}^*\over V_{ts}^*} {M_W^2 \over 
            M_{sq}^2} {1\over \sw} \sum_{i,j=1,2} (V_{i1}^* V_{j1} V_{i1} V_{j1}^* )
            f(x_i,x_j,x_{\tilde \nu}) \ , 
        \end{equation}
        where:
        \begin{equation}
            f(x_i,x_j,x_{\tilde \nu}) = {1\over 2}\int_0^1 dx \int_0^1 dy \int_0^1 dz 
            { yz(1-z)^2  \over \left[ y(1-z) + x_{\tilde \nu} (1-y)(1-z) +z (x_i
            x+x_j (1-x)) \right]^2} \ ,
        \end{equation}
        and $x_{\tilde \nu} = M^2_{\tilde \nu} / M^2_{sq}$. \\
        \\
\textbf{Box diagram with higgsino-bottom-stop vertex: } \\ 
\\
Replacing the wino with a higgsino yields
        \begin{equation}
            C_9 = - C_{10} = -(\delta^u_{23})_{LR} {V_{cs}^*\over V_{ts}^*} {M_W^2 \over 
            M_{sq}^2} {\lambda_t \over g_2 \sw} \sum_{i,j=1,2} (V_{i1}^* V_{j1} 
            V_{i1} V_{j2}^*) 
           f(x_i,x_j,x_{\tilde \nu}) \ .
        \end{equation}

\section{Feynman integrals and hypergeometric functions}\label{sec:hypgeo}
 
\subsection{Hypergeometric functions and integral representations}
 To calculate the new Wilson coefficients in the NMFV/MIA framework, the following integrals need to be evaluated :
\begin{equation}
P_{ijk}(a,b)\equiv \int_0^1 d x \int_0^1 d y \frac{y^i (1-y)^j}{(1 - y + a x y  + b (1-x) y)^k} \ .
 \label{eq:def_of_Pijk}
\end{equation}
These integrals can be shown to be linear combinations of hypergeometric functions $_p F_q$ (for an extensive review see Refs.~\cite{bateman1953higher,slater2008generalized}). In most cases, the integrals can be rewritten using solely $_2 F_1$, which is sometimes referred to as the Gaussian or ordinary hypergeometric function. This leads to nearly arbitrary precision in the numerical implementation as the series usually converge quickly. However, in some cases, particular analytical continuation formulae have to be used, which are well known in the literature, and can be found in e.g. Ref.~\cite{bateman1953higher}.

The integral representation of $_2 F_1$ is defined by Euler's formula:
\begin{equation}\label{eq:hypegeo_def}
_2F_1(a,b;c;z)  = \frac{\Gamma(c)}{\Gamma(b)\Gamma(c-b)} \int_0^1 t^{b-1}(1-t)^{c-b-1}(1-tz)^{-a}\, \dd t \ , 
\end{equation}
which is a one-valued analytic function of $z$, provided $\Re(c) > \Re(b) > 0 , \abs{\arg(1-z)} < \pi $\cite{bateman1953higher}. \\
In what follows, we will simply refer to $_2F_1$ as $F$. 
Equation~(\ref{eq:hypegeo_def}) provides an analytic continuation of $F$, which is usually defined by its series representation: 
\begin{equation}\label{hypseries}
    F(a,b;c;z) := \sum_{n = 0}^{\infty} \frac{(a)_n (b)_n}{(c)_n} \frac{z^n}{n!} \ ,
\end{equation}
where : 
\begin{equation*}
    (a)_n = \frac{\Gamma(a+n)}{\Gamma(a)}, \textrm{ is the (rising) Pochhammer symbol} .
\end{equation*}

The generalized hypergeometric functions $_p F_q$ can be defined by
\begin{equation}\label{eq:pFq}
    _pF_q(\left[ 
\begin{array}{c}
a_{1},\ldots ,a_{p} \\ 
b_{1},\ldots ,b_{q}
\end{array}
;z\right] = \sum_{n=0}^\infty \frac{(a_1)_n\cdots(a_p)_n}{(b_1)_n\cdots(b_q)_n} \, \frac {z^n} {n!} \ ,
\end{equation}
and a general Euler integral transform relates hypergeometric functions of higher and lower orders~\cite{slater2008generalized}:
\begin{equation}\label{eq:general_euler_transform}
 _{A+1}F_{B+1}\left[ 
\begin{array}{c}
a_{1},\ldots ,a_{A},c \\ 
b_{1},\ldots ,b_{B},d
\end{array}
;z\right] =\frac{\Gamma (d)}{\Gamma (c)\Gamma (d-c)}
\int_{0}^{1}t^{c-1}(1-t)_{{}}^{d-c-1}\ {}_{A}F_{B}\left[ 
\begin{array}{c}
a_{1},\ldots ,a_{A} \\ 
b_{1},\ldots ,b_{B}
\end{array} ; tz\right] \ .
\end{equation}

\subsection{Analytical expressions for $P_{ijk}$ }
Let us demonstrate that the class of loop integrals in (\ref{eq:def_of_Pijk}) can all be expressed using hypergeometric functions. 
\paragraph{}
First, we rewrite \eqref{eq:def_of_Pijk} using Fubini's theorem as
\begin{align}
    P_{ijk}(a,b) &= \int_0^1 \dd y \, y^i (1-y)^j \int_0^1 \dd x \frac{1}{\left[ A(a,b;y)x + B(b;y)\right]^{k}} \label{eq:integral} \ ,\\
    \shortintertext{with: }
    A(a,b;y) &= y(a-b) \ , \\
    B(b;y) &= 1 + y(b-1) \ .
\end{align}
Therefore, for $\mathbf{k \neq 1}$ and $\mathbf{a \neq b}$, we can integrate over $x$ using Cavalieri's quadrature formula:
\begin{equation}\label{eq:Cavalieri}
    \int_0^1 \ \frac{\dd x}{(B + Ax)^k} = \left[\frac{(Ax + B)^{1-k}}{(1-k)A} \right]^1_0 = \frac{(A+B)^{1-k} - B^{1-k}}{A(1-k)} \ .
\end{equation}
Then, by replacing the integrand in \eqref{eq:integral} with the result of \eqref{eq:Cavalieri} we obtain

\begin{equation}
    P_{ijk}(a,b) = \int_0^1 \dd y \, \frac{y^{i-1}(1-y)^j}{(a-b)(1-k)} \,
    \left[ (y(a-1) +1)^{1-k} - (y(b-1) +1)^{1-k} \right] \;.
\end{equation}
We can spot the integral representation of the hypergeometric function. Explicitly:
\begin{equation}
\begin{aligned}
    P_{ijk}(a,b) &= \frac{1}{a-b} \frac{1}{1-k}
      \int_0^1 \dd y \, y^{i-1} (1-y)^j \left[
      (1 - y(1-a))^{1-k} - a \leftrightarrow b 
      \right] \\
    &= \frac{ \beta_E(i,j+1)}{(a-b)(1-k)}
    \left[
    F(k-1,i;i+j+1;1-a)- a \leftrightarrow b 
    \right] \ ,
\end{aligned}
\end{equation}
where we use Eq. (\ref{eq:hypegeo_def}) and the Euler's beta function's definition.

\underline{$\mathbf{k \neq 1 \;,\; a = b} $}: \\ \\
Let $G(b) = F(k-1,i;i+j+1;1-b)$. By definition: 

\begin{align}
    \lim_{a\rightarrow b} \frac{G(a) - G(b)}{a - b} &= G'(b) \ , \\
    \intertext{and the differentiation formula for $F$ is:}
    \dv{F(\alpha,\beta;\gamma;z)}{z} &= \frac{\alpha \beta}{\gamma} F(\alpha+1,\beta+1; \gamma+1; z) \ , \\
    \intertext{therefore:}
    \lim_{a \rightarrow b} P_{ijk}(a,b) &= \beta_E(i,j+1) \lim_{a \rightarrow b} \; \frac{1}{(1-k)(a-b)}  \Big[G(a) - G(b)\Big] \ , \\
    &= \fr{\beta_E(i,j+1) }{1-k} \, G'(b)  \ ,\\
    &= \fr{1}{1-k} \beta_E(i,j+1) \, \fr{(k-1)i}{(i+j+1)} 
    F(k, i+1; i+j+2; 1-b) \ ,\\
    \intertext{which simplifies to : }
    P_{ijk}(b,b) &= \beta_E(i+1,j+1) \, F(k,i+1,i+j+2; 1-b) \ .
\end{align}

\underline{$\mathbf{k = 1}$:}\\ \\
Only three of these integrals appear in the computation of the Wilson coefficients, all three are reasonably doable:
\begin{align}
    P_{111}(a,b) &= \int_{[0,1]^2} \dd{x} \dd{y} \frac{y(1-y)}{1 - y + axy + b(1-x)y} \ ,\\
    P_{021}(a,a) &= \int_{[0,1]^2} \dd{x} \dd{y} \frac{(1-y)^2}{1 + y(a-1)}  \ ,\\ 
    P_{111}(a,a) &= \int_{[0,1]^2} \dd{x} \dd{y} \frac{y(1-y)}{1 + y(a-1)} \ .
\end{align}
The last two integrals are straightforward to compute: 
\begin{align}
    P_{111}(a,a) &= \int_0^1  \dd y \; \frac{y(1-y)}{1 + y(a-1)}  \ , \nonumber \\
    &= \fr{1}{a-1} \Bigg(
   \Big[ y(y-1)\log{}(1+ y(a-1)) \Big]^{1}_{0} - \int_0^1 (1-2y)\log{}((1+y(a-1)) \dd{y} 
    \Bigg)  \ , \nonumber \\
    &= \fr{-1}{a-1} \int_0^1 \dd{y} (1-2y)\log{}((1+y(a-1))  \ ,\nonumber \\ 
    &= \fr{-1}{a-1} \frac{-a^2 + 2a \log{(a)} + 1}{2(a-1)^2}  \ ,\\
    \shortintertext{with:}
    &\lim_{a \rightarrow 1}  \fr{-1}{a-1} \frac{-a^2 + 2a \log{(a)} + 1}{2(a-1)^2} = \fr{1}{6} \ . \nonumber
\end{align}
Similarly,
\begin{align}
    P_{021}(a,a)&= \int_{[0,1]^2} \dd{x} \dd{y} \frac{(1-y)^2}{1 + y(a-1)} = \frac{-1}{a-1} \int_0^1 \dd{y} 2(1-y)\log{(1+y(a-1))} \ , \nonumber  \\
    &= \frac{(a+1)(3a^2 - 2a^2 \log{(a)}- 4a+1}{2(a-1)^2 a} \ ,  \\
    \shortintertext{with:}
    &\lim_{a \rightarrow 1} \frac{(a+1)(3a^2 - 2a^2 \log{(a)}-4a+1}{2(a-1)^2 a} = 0  \ ,\nonumber 
\end{align}

\begin{align}
    P_{111}(a,b) &= \int_{[0,1]^2} \dd{x} \dd{y} \frac{y(1-y)}{1 - y + axy + b(1-x)y} \ ,\\
    &= \fr{1}{2(a-1)^2(a-b)(b-1)^2(ab-a-b+1)} \times \\
    &\Big[
    (a - 1)^{2}(a - b)(b - 1)^{2} + (a - 1)^{2} (b - 1)^{2} \log{\left(\fr{a}{b}\right)}(a b - a - b + 1)  \nonumber \\
    &+(a - 1)^{2} (2 b - 1) (-\log{b})(a b - a - b + 1) \nonumber \\
    &+(2 a - 1) (b - 1)^{2} \log{(a)}(a b - a - b + 1) \Big] \nonumber \ .
\end{align}

\paragraph{}
For completeness, let us show that in this case too, we can re-express $P_{ij1}(a,b)$ using hypergeometric functions:\\
We are interested in the set of integrals defined by
\begin{equation}\label{eq:Pij1}
    P_{ij1}(a,b) = \int^1_0 \dd{x} \int^1_0 \dd{y} \frac{y^i(1-y)^j}{1 - y + axy + b(1-x)y} \ .
\end{equation}
Two cases, $a=b$ and $a\neq b$, have to be distinguished

\underline{$\mathbf{k=1, a = b}$:}\\ \\
In this case, the computation is straightforward and yields

\begin{align}
     P_{ij1}(a,b) &= \int^1_0 \dd{x} \int^1_0 \dd{y} \frac{y^i(1-y)^j}{1 - y + axy + b(1-x)y}\ , \\
        &= \int^1_0 \dd{y} \fr{y^i (1-y)^j}{(1-y(1-a))}\ , \nonumber \\
        &= \beta_E(i+1,j+1) _2F_1(1,i+1;i+j+2;1-a)\ .
\end{align}

\underline{$\mathbf{k=1, a \neq b}$:}\\ \\
We can express Eq. (\ref{eq:Pij1}) as an integral over $_2F_1$:
\begin{align}
P_{ij1}(a,b) &= \int^1_0 \dd{x} \int^1_0 \dd{y} \frac{y^i(1-y)^j}{1 - y + axy + b(1-x)y} \ ,\\
        &= \int^1_0 \dd{x} \int^1_0 \dd{y} y^i(1-y)^j (1-y(1-b-x(a-b))) \ , \nonumber \\
        &= \int^1_0 \dd{x} \beta_E(i+1,j+1) _2F_1(1,i+1;i+j+2;1-b-x(a-b)) \ .
\end{align}
This holds provided $\Re{1-ax-b(1-x)} < 1$ . For $x \in [0,1], a,b \in \mathcal{R}^+$, this is always true.\\
Using the general Euler transform (\ref{eq:general_euler_transform}) we can compute the integral over $_2F_1$:

\begin{equation}
    P_{ij1}(a,b) = \beta_E(i+1,j+1) \frac{1}{a-b} \Bigg( (1-a) \, _3F_2 \left[\begin{array}{c}
         
         1,1,i+1  \\
         2,i+j+2 
    \end{array};1-a\right] - a \leftrightarrow b\Bigg) \ .
\end{equation}
    
The two previous results for $a=b$ and $a\neq b$ can be checked against the explicit analytical expressions given before.

\bibliographystyle{jhep.bst}
\bibliography{biblio}

\end{document}